\def\lesssim{{_ <\atop{^\sim}}}

\def\ap3m{AP$^3$M}

\def\Msun{${\rm M_{\odot}}$}
\def\kms{$\,{\rm{\ }km{\ }s^{-1}}$~}

\def\lesssim{\mathrel{\hbox{\rlap{\hbox{\lower4pt\hbox{$\sim$}}}\hbox{$<$}}}}
\def\gtrsim{\mathrel{\hbox{\rlap{\hbox{\lower4pt\hbox{$\sim$}}}\hbox{$>$}}}}

\documentstyle[epsfig]{mn}

\begin{document}

\title{Exploring spiral galaxy potentials with hydrodynamical simulations}
\author[Slyz A. et al.]
       {Adrianne D. Slyz$^{1}$, Thilo Kranz$^{2,3}$, 
        and Hans-Walter Rix$^{2}$\\        
       {$^1$University of Oxford, Astrophysics, Denys Wilkinson Building, Keble Road, Oxford OX1 3RH, United Kingdom}\\
       {$^2$Max-Planck-Institut f\"{u}r Astronomie, K\"{o}nigstuhl 17, Heidelberg 69117, Germany}\\
       {$^3$current address: German Aerospace Center DLR, K\"{o}nigswinterer Strasse 522-524, 53227 Bonn, Germany}}

\date{Received ...; accepted ...}

\maketitle

\begin{abstract}
We study how well the complex gas velocity fields induced by massive
spiral arms are modelled by the hydrodynamical simulations
we used to constrain the dark matter fraction 
in nearby spiral galaxies (Kranz et al. 2001, 2003) . 
More specifically, we explore the dependence of the positions 
and amplitudes of features in the gas flow on  
the temperature of the interstellar medium (assumed to behave
as a one-component isothermal fluid), the non-axisymmetric disk contribution 
to the galactic potential, the pattern speed, $\Omega_{\rm p}$ and finally
the numerical resolution of the simulation.
We argue that, after constraining the pattern speed reasonably well by 
matching the simulations to the observed spiral arm morphology, the amplitude 
of the non-axisymmetric perturbation (the disk fraction) is left as the primary
parameter determining the gas dynamics. However, due to the  
sensitivity of the positions of the shocks to modeling 
parameters, one has to be cautious when quantitatively
comparing the simulations to observations.
In particular, we show that a global least squares analysis is not the
optimal method for distinguishing different models as it tends to slightly favor
low disk fraction models.
Nevertheless, we conclude that, given 
observational data of reasonably high 
spatial resolution and an accurate shock-resolving hydro-code 
this method tightly constrains the dark
matter content within spiral galaxies. We further argue that even if 
the perturbations induced by spiral arms are weaker than those of 
strong bars, they are better suited for this kind of analysis
because the spiral arms extend to larger radii 
where effects like inflows due to numerical viscosity and morphological dependence
on gas sound speed are less of a concern than they are in the centers of disks.

\end{abstract}


\section{Introduction} \label{introduction}
Because gas responds strongly to nonaxisymmetries
in a gravitational field, it was recognized more than two
decades ago as a sensitive tracer of galactic potentials.
Therefore, a model for such a potential can be 
tested by simulating the gas flow within it, and 
comparing the resulting morphology and kinematics to observations.
The earliest efforts to apply such a method
used general forms for the potential
derived either from N-body simulations (Huntley 1978) or from
analytic considerations (Sanders and Tubbs 1980).
The parameters of these model potentials were then
constrained by comparing results from hydrodynamical simulations performed with
the beam scheme (Sanders and Prendergast 1974) to the 
morphology and kinematics of NGC 5383. The aim was to
understand how the general
features of the gas in a typical SBb(s) galaxy arose.
Since NGC 5383 was being used as a representative of SBb(s) galaxies, Duval and
Athanassoula (1983) recognized the importance of doing a careful observational study
of it and hence obtained more complete spectral and photometric
data for it.  However using better
data did not resolve the discrepancies between modeled  and observed kinematics. 
They blamed it both on an inhomogeneity of the observations and an inadequacy
of the models. Subsequent efforts to constrain
disk galaxy potentials via hydrodynamical simulations
have benefited from improvements
in hydrocodes and have focused on deriving galactic potentials from
specific galaxies rather than assuming a general form for them
(England 1989, Garcia-Burillo, Combes \& Gerin 1993,
Sempere, Garcia-Burillo, Combes, \& Knapen 1995, Sempere, Combes, \& Casoli 1995,
Lindblad, Lindblad and Athanassoula 1996, 
Lindblad and Kristen 1996, Sempere \& Rozas 1997).

Interest in this approach has recently resurfaced with the defined goal
of discerning the dark matter content of disk galaxies. 
For the case of barred galaxies, where gas motion in the
inner region is strongly non-circular,  with velocity gradients on the order of
hundreds of kilometers a second, Weiner, Sellwood \& Williams 2001
argued that model fits to the observed velocity
fields could unequivocally differentiate between maximal and
submaximal disks.
With the same goal in mind, we undertook an investigation to break the 
disk-halo degeneracy in any spiral galaxy by studying
the response of gas flow to the  weaker perturbations induced
by spiral arms. We began the study with 
the spiral galaxy NGC 4254 (Kranz, Slyz \& Rix (2001, hereafter Paper I)
and then applied the method to a sample of four additional high-surface brightness,
late-type spiral galaxies (NGC 3810, NGC 3893, NGC 5676 and NGC 6643) 
(Kranz, Slyz \& Rix 2003).
To summarize, the studies were based upon the assumption that provided
the dark matter halo is axisymmetric, all non-axisymmetric features observed in
the velocity field of the gaseous disk
 have to be generated by the stellar mass component.
Therefore, while completely smooth rotation curves do not
betray any information about the dark and baryonic fractions
of galaxies, non-axisymmetric features in rotation curves
might break the baryonic/dark matter degeneracy.

We expected that when gas in our simulations would cross a spiral
feature in the potential there would be a ``wiggle'' in the velocity
field whose amplitude would
be proportional to the local stellar mass fraction.
Ideally, the simulated velocity wiggles would correlate well with those
measured in the observed gas velocity field if the 
gravitational potential used for the simulations was derived from the observed
mass distribution.
However in modelling the gas flow in different galaxies we found that
we could not account for every feature observed in the
H$\alpha$ gas kinematics.
The identifications of kinematical features in the simulations with
those in the observations were often ambiguous. Kinematical features in the
simulations sometimes appeared to be displaced, and/or to have a
different profile, or amplitude.

Before considering the addition of more physics to our simulations,
such as self-gravity or a multiphase interstellar medium sustained 
by star formation, supernovae and stellar winds,
we propose in this paper to take a closer look at how the gas flow
in a model of one of our sample galaxies, NGC 4254, changes as a function of 
our model parameters. More specifically, we investigate how much of the 
mismatch that we see is due to the coarseness of our parameter space 
exploration and how sensitive the positions, amplitudes, and profiles 
of features in the gas flow are to the parameters.
In parallel, we also search for systematic measurements that can gauge the 
accuracy of a model potential.
This allows us to estimate the numerical error associated 
with our simple hydrodynamical model, and therefore to
assess the robustness of our result for the disk fraction of
NGC 4254, namely that it is $\leq$ 85 per cent
of its maximal value, implying
that $\geq$ 1/3 of the total mass within 2.2 $K$-band disk scale 
length is dark (Paper I).

Hence, the paper is organized as follows.
Section~\ref{hydro_modeling}  gives a brief description of the 
hydrodynamical method we use and it introduces the parameter space
we explore in this paper. It further describes the initial and boundary
conditions of our models, addresses
the question of whether the gas flow in our simulations reaches
a quasi steady-state, explores the 
choice of the initial gas density profile and grid resolution.
Section~\ref{Parameters} proceeds to examine how
the gas flow in the spiral
potential depends on the modeling parameters, namely
the contribution of the non-axisymmetric
component to the galaxy potential, the gas sound speed and the 
potential's pattern speed. Section~\ref{massinflow} deals with the
mass inflow as a function of the examined parameters and 
the hydrodynamical method. Section~\ref{discussion} gives our interpretation of 
the parameter study based on two different approaches to determine the 
quality of match between observations and simulations. Finally
section~\ref{conclusions} presents our conclusions.

\section{Hydrodynamical Modeling} \label{hydro_modeling}

Our simulations are carried out using the BGK (Bhatnagar-Gross-Krook)
hydrocode, a code based on gas-kinetic theory (Prendergast and
Xu 1993, 
 Slyz and Prendergast 1999). This is a high-resolution,
Eulerian, grid-based hydrodynamics code.
At each grid wall,
BGK computes time-dependent hydrodynamical fluxes from velocity
moments of a distribution function which is a local solution
to a model of the collisional Boltzmann equation, namely the
BGK equation. Because the
BGK scheme evolves gas flow through an equation that includes
particle collisions, the fundamental mechanism for generating
dissipation in gas flow, the BGK flux expressions carry both 
advective and dissipative terms. If the grid is not fine enough
to resolve a shock, then the collision time which is recomputed
at each wall of the grid and with each timestep is
enlarged to increase the viscosity and heat conduction at that
particular location. Thereby 
even when the dissipation is
put in for numerical reasons, it is added into the fluxes in exactly 
the same way that the physical dissipation is put into the code, hence there is no
source term for either the physical or artificial dissipation.
The code has been extensively
tested on standard 1D and 2D test cases of discontinuous
nonequilibrium flows (see Xu 1998 for a review). It has been
used to solve Navier-Stokes problems in smooth flow regions
both with (Slyz et al. 2002) and without (Xu and Prendergast 1994)
gravity, and it has been tested for its long-term stability 
and convergence to the equilibrium solution in a fixed 
external gravitational field (Slyz and Prendergast 1999). 

One reason for carrying out the disk simulations with this 
code, is its low diffusivity, a property that is critical not 
only to capture the shocks that form when the gas orbits in
the non-axisymmetric potential, but also to properly model the 
loss of angular momentum and hence
the resulting radial inflow of the gas due to the strong shear 
in the underlying differentially
rotating disk. Slyz et al. (2002) showed that if an isothermal
gas is initialized to be in centrifugal equilibrium within a 
purely axisymmetric galactic potential, simulation with 
the BGK scheme produces the steady-state Navier-Stokes solution 
to a high degree of accuracy. The tests were
carried out for parameters which are relevant for
galaxy studies: an asymptotically flat rotation curve with
$v_{\rm max} = $ 220 \kms, a sound speed of $c_{\rm s} = $ 10 \kms, i.e.
a highly supersonic (Mach $\approx$ 20) shear flow throughout
most of the disk. The success of BGK in giving viscous radial 
flows on the order of 1 \kms in a disk rotating differentially at 
220 \kms is a technical success which 
insures that when studying the kinematics of the gas in a galactic
disk, with a decent grid resolution, one does not have to worry
about artificial dissipation.

The number of grid cells, i.e. the spatial resolution of a simulation, 
is one of the parameters whose variation we study in this paper.
In addition to the dissipation introduced by the BGK algorithm, there is
the inevitable dissipation arising from the fact that the code only saves cell
averages at the end of each iteration. Hence the larger the cells, the
less information the code retains. To keep this numerical dissipation
which is proportional to the cell dimensions
at a constant value throughout the grid, we  
perform our simulations on an evenly spaced Cartesian grid.
Our runs in Paper I were performed on a 201 $\times$ 201 grid
giving a resolution element of about 115 pc on a side. For comparison,
in this paper we look at runs done at half 
(101 $\times$ 101) and double that resolution (401 $\times$ 401).

There are three other parameters we explore in our modeling.
As already stated in the introduction and described in Paper I, 
for our numerical investigation of the solutions for gas flow in
the gravitational potential of NGC 4254 we use a potential derived
from observations.  The mass-to-light ratio corrected $K$-band image provides
us with a stellar density map from which we compute the form of the
non-axisymmetric component of the gravitational potential, and the
rotation curves from observed long-slit H-alpha kinematics
give us a measurement of the total gravitational potential of the galaxy. 
By assuming an axisymmetric isothermal profile for the
dark halo we construct a series of potentials of different values for the
strength of the stellar contribution, $f_{\rm d}$ (cf. eqns. 9 and 10, Paper I),  
which all match the observed rotation curve. 
Constraining the parameter $f_{\rm d}$ is our main scientific objective. 
Note that we do not work with a self-consistent model.
We dynamically follow the gas, neglecting its
self-gravity, in a fixed external potential which represents the combined
gravitational effect of stars and dark matter.

\begin{figure}
 \centerline{\psfig{file=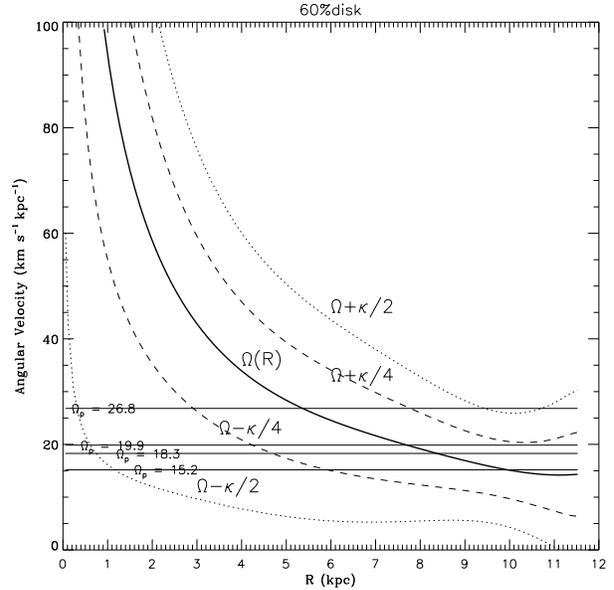,width=\hsize}}
 \caption{Plot of NGC 4254's angular velocity $\Omega(R)$ (thick solid line)
 derived from an axisymmetric fit to its rotation curve.
 Overplotted are the angular velocities, $\Omega_{\rm p}$, of the 
 different rigidly rotating reference frames which we
 explore (solid horizontal lines), and curves showing $\Omega - \kappa/m$ where m 
 is 2 (dotted curves) and 4 (dashed curves).}
 \label{resonances}
\end{figure}

Another parameter which plays the most important role
in shaping the gas morphology in our simulations is the pattern speed, 
$\Omega_{\rm p}$, of the external potential. We assume the entire potential rotates
rigidly with the same time-independent pattern speed and we
perform the simulations in this rotating reference
frame. We choose the direction of pattern and gas rotation to be clockwise
so that inside corotation the gas enters the spiral arms from the concave side.
Figure~\ref{resonances} shows how the locations of the resonances
change with the different $\Omega_{\rm p}$ that we use.

We keep away from the difficult question of how the
spiral formed, and we do not look for time-dependent solutions. Instead we study
only steady or quasi-steady flows in 
NGC 4254's fixed external gravitational potential.
In the time-independent case, the gas flow must satisfy:
\begin{equation}
\vec{u} \cdot \nabla \vec{u} + 2 \Omega_{\rm p} \times \vec{u} = -\nabla p/\rho \pm \nabla \Phi
\end{equation}
\begin{equation}
\nabla \cdot (\rho \vec{u}) = 0
\end{equation}
where $\Phi$ is the potential of the combined centrifugal and gravitational
forces. This system of equations must be completed by
an equation of state and this introduces the last parameter
of the problem: the gas sound speed. Because we
do not know the effective equation of state of interstellar matter,
the simplest thing to do is to assume an isothermal equation
of state $p = K \rho$, where {\em K} 
$= {c_{\rm s}}^{2}/\gamma$,  $\gamma$ being the ratio of specific
heats of the gas, and $c_{\rm s}$ is its constant sound speed. To study how
the gas flow responds to changes in ${c_{\rm s}}$, we simulate the
gas with sound speeds of 10, 15, 20 and 30 \kms.

Since the scale height of gas and stars in a typical non-interacting late type disk galaxy
is about 1/40 to 1/75 the diameter of the visible galactic disk (e.g., Schwarzkopf \& Dettmar 2000), we restrain this study to two-dimensions.
More specifically we approximate the disk as a thin sheet, and only compute
the gas flow in the two dimensions of the disk plane. 
Alternatively, one can view this approximation as an integration
over the disk thickness perpendicular
to the plane, and our physical variables as mean values in this direction.

Table~\ref{SimuParam} summarizes the parameters we use for
the simulations and indicates the parameters of our fiducial simulation in 
boldface.
 
\subsection{Initial conditions}
We initialize the gas density profile to be exponential
with a scale length which is on the order of the
observed disk's stellar scale length, namely 3.86 kpc. 
Upon estimating the
total mass of the galaxy from the observed rotation curves,
we set the mass of the gaseous disk to be 5 per cent of this
total mass. The gas is therefore moving in a potential
produced by a mass much greater than itself which means that,
even in the densest regions (spiral arms), the neglect of its 
self-gravity will translate into a modest underestimate of its density
(Berman, Pollard and Hockney 1979).

\begin{table}
\caption{Specifications of the hydrodynamical simulations. Values for fiducial run given in boldface.}
\label{SimuParam}
 \begin{tabular}{ll} 
  Simulation Parameters \\\hline
  grid length (kpc)          & 23.2 \\
  $\rm{v}_{\rm max}$ (\kms)           & 152  \\
  Initial Gas Mass     & $3.14 \cdot 10^{9}$\Msun \\  
  number of grid cells & $101^2$,${\bf201^2}$,$401^2$ \\
  time to turn on full potential ($\rm{t}_{\rm FP}$)& $20{\ } \rm{t}_{\rm s}$,{\bf$40{\ } \rm{t}_{\rm s}$},$80{\ } \rm{t}_{\rm s}$\\
  time for entire simulation    & $\rm{t}_{\rm FP} + 2{\ }\rm{t}_{\rm dyn}$ \\
  gas sound speed (\kms)     & {\bf10}, 15, 20, 30  \\
  $\Omega_{\rm p}$ (\kms/kpc)          & 26.8, {\bf19.9}, 18.3, 15.2  \\
  ${\rm R}_{\rm c}$ (kpc) (corotation radius)     &  5.6, {\bf7.58}, 8.5, 10  \\
  ${\rm f_{\rm d}}$ disk fraction (\%) & 20, 44, {\bf60}, 85, 100 \\
 \end{tabular}
\end{table}

\begin{figure*}
 \centerline{\psfig{file=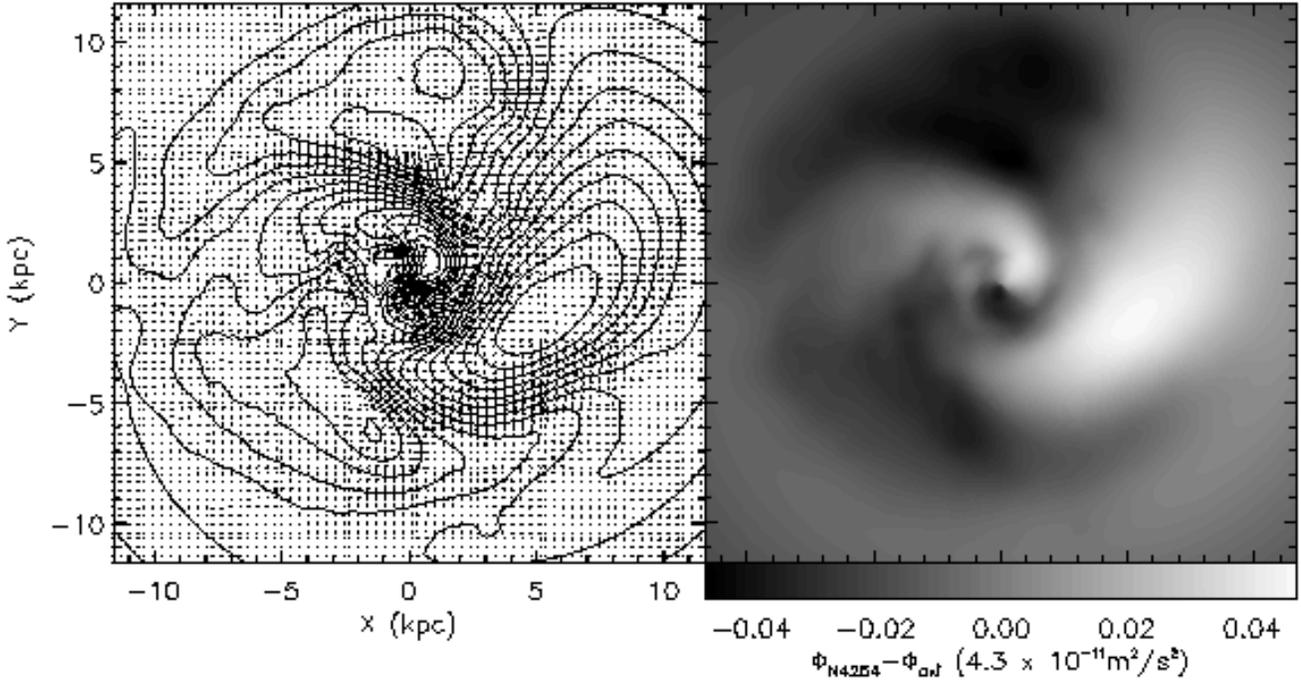,width=\hsize}}
 \caption{The axisymmetric potential subtracted from the potential of
 NGC 4254.  Contours are shown on the left with the force vectors,
 computed from the difference in the two potentials, overplotted.
 A filled contour plot is shown on the right, with the peaks of the
potential given in white and the valleys in black.}
 \label{Potential}
\end{figure*}

As for the initial dynamics of the gaseous disk, the simulations begin 
with the gas flowing on circular orbits
in inviscid centrifugal equilibrium with respect to the 
axisymmetric gravitational potential which best fits the 
observed rotation curves. The non-axisymmetric perturbations 
are then gradually turned on to avoid transient structures (Sorensen \& Matsuda 
1982). 
The criteria for the time in which the full potential
is turned on, ($t_{\rm FP}$), is based on the sound crossing
time across the diagonal of a grid cell ($t_{\rm s}$). 
For the 201 $\times$ 201 grid, we set $t_{\rm FP} = 40{\ }t_{\rm s}$.
Since the sound crossing time depends on the length of
the grid cell, and for comparison's sake we want $t_{\rm FP}$ and the total
running time of each simulation to be identical, 
this implies $t_{\rm FP} = 20{\ }t_{\rm s}$
for the 101 $\times$ 101 grid, 
and $t_{\rm FP} = 80{\ }t_{\rm s}$ for the 401 $\times$ 401 grid.
In terms of the dynamical time of the outer edge of the disk
(${\rm t}_{\rm {dyn}} = \frac{2 \pi R_{\rm {disk}}}{V_{c}} \approx$ 480 Myr 
for a rotational velocity
of about 152 \kms at ${\rm R}_{\rm {disk}} = 11.6$ kpc), for
the case where the sound speed is 10 \kms and the grid is 201 $\times$ 201, 
this means that we turn
on the full potential in 1.3 ${\ } {\rm t}_{\rm {dyn}}$. 
After the full potential is turned on, we continue to run the
simulations for another two dynamical times.
Figure~\ref{Potential} shows the non-axisymmetric component of the
gravitational potential of NGC 4254 displayed on the 201 $\times$ 201 grid.

\begin{figure*}
 \centerline{\psfig{file=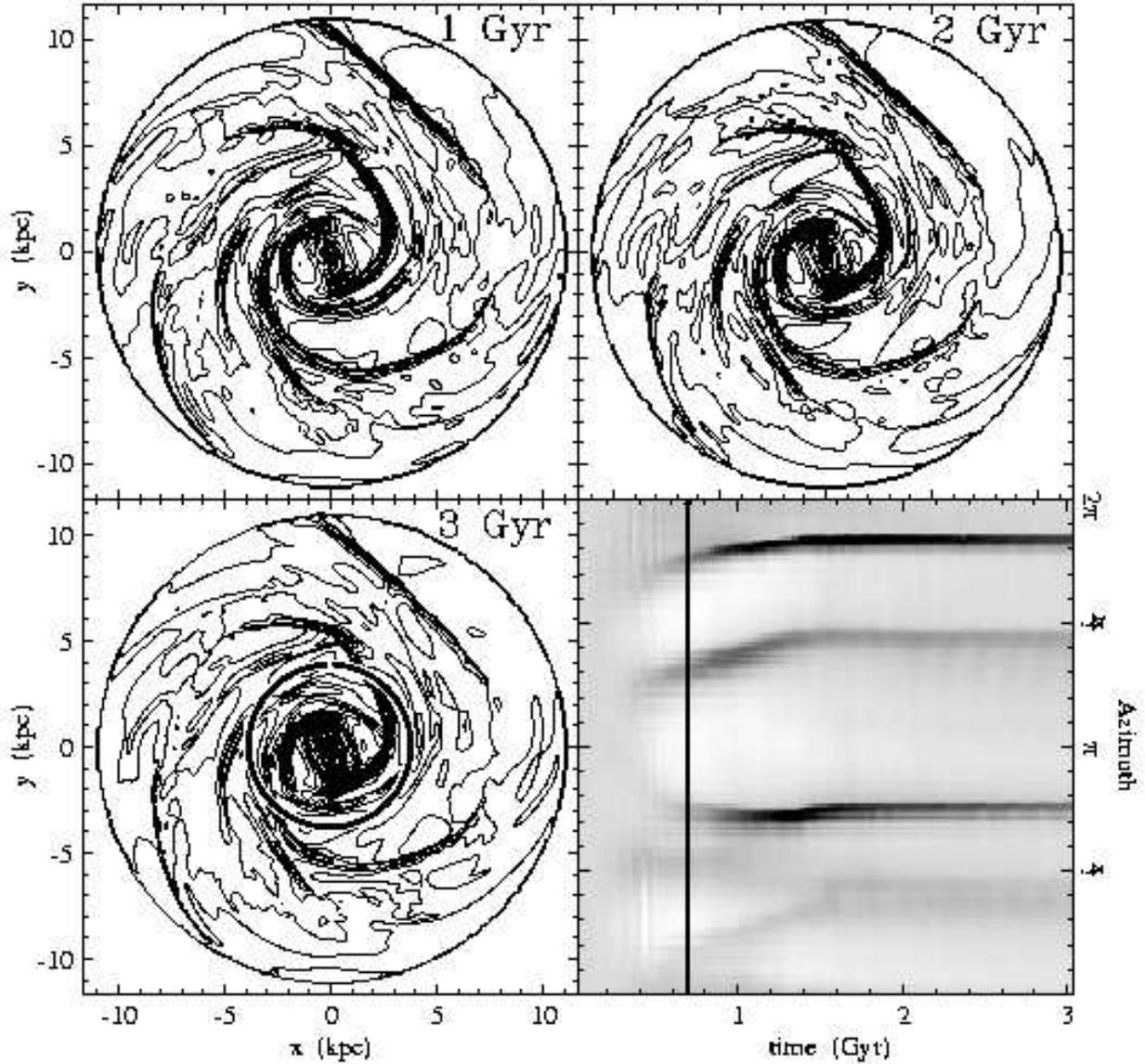,width=\hsize}}
 \caption{Contour maps showing three time steps separated by a Gyr
 of the simulated gas density for our fiducial run 
 in order to
 demonstrate that the gas flow reaches a steady-state. The shaded plot in the
 lower right hand corner shows the density as a function of the
 azimuth along a circle of radius $R = 3.8 \,\rm{kpc}$ (indicated by a thick
 solid line on the contour plot at 3 Gyr) as a function of time. The
 solid vertical line in this plot indicates the time, $\rm{t_{\rm FP}}$,
 at which the full
 potential is turned on in the simulation. After $\rm{t_{\rm FP}} +\rm{t_{\rm dyn}}$ ($\approx 1.1$ Gyrs) the simulation yields a very
 stable pattern.}
 \label{Longevol758}
\end{figure*}
   \begin{figure}
      \centerline{\psfig{file=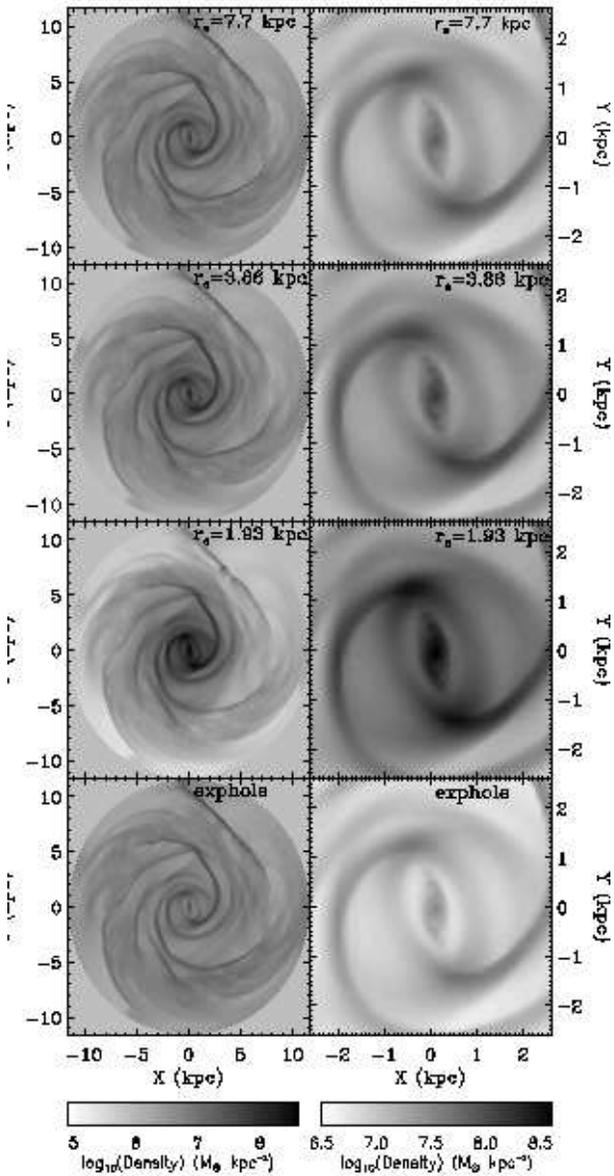,width=.95\hsize}}
      \caption{Grey scaled maps of the log of the density from
simulations with our fiducial parameters
and with different initial gas density
profiles. The entire simulated region is shown in the left column
      and the inner 2.6 kpc$^2$ region is shown in the right column.}
      \label{morph_diffICS}
    \end{figure}
   \begin{figure}
      \centerline{\psfig{file=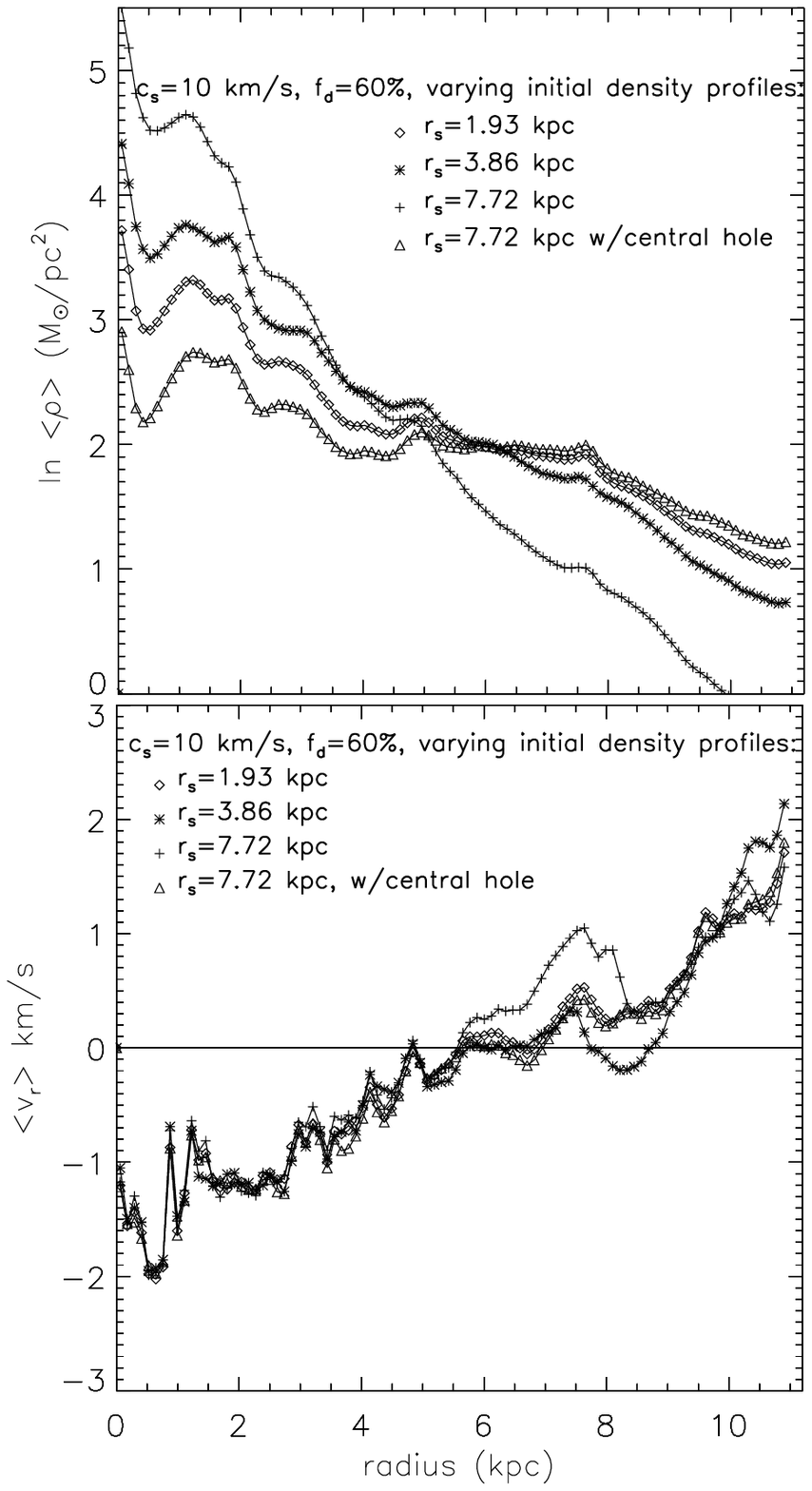,width=.95\hsize}}
      \caption{Natural logarithm of the average density (top panel)
      and density averaged radial
velocity (bottom panel) at time $t = \rm{t}_{\rm FP} + 2{\ }\rm{t}_{\rm dyn} \approx $1.6 Gyr for runs differing only
in their initial density profile. The other parameters in the
simulation are the fiducial ones.}
      \label{diffICS_dens_rvel}
    \end{figure}

\subsection{Boundary conditions}

Since we perform our computations on a Cartesian grid, the center of 
the disk (r = 0) is not a singular point, and therefore does 
not require special treatment via an inner boundary condition. 
Instead, the gas flow is computed through this point exactly as it is computed 
throughout the grid.

An outer boundary condition is, however, unavoidable.
To tackle this issue, we keep two ``rings''
of one cell thick ghost cells outside of a radius of $R_{\rm disk}$. Beyond
these ghost cells we do not follow the evolution of the
gas. Hence we have effectively carved a circular grid out
of the square Cartesian grid. At the end of each simulation timestep
we update the values of the hydrodynamic quantities (mass,
momentum and energy) in the ghost cells by performing a
bilinear interpolation to the cells in the vicinity of
the ghost cell. To be more specific, for each ghost cell
in the inner ring for example, we compute the coordinates of
the intersection of the line extending radially from the
center of the disk to the ghost cell with the circle bounding
the true flow region of the grid. We then find the four cells
surrounding this intersection (some of which might be other
ghost cells). After fitting a surface to the hydrodynamic
quantities in these four cells we assign the ghost cell
the value the fitted surface has at the intersection. 
By filling up the ghost cells via constant radial extrapolation
that varies azimuthally around the disk, we are better able to
handle situations in which there is a significant non-axisymmetry
near the outer boundaries. For example, from the map of NGC 4254's 
potential (Figure~\ref{Potential}), one can see that the potential
is quite non-axisymmetric near the upper boundaries thereby
requiring an outer boundary condition which can take into
account the possiblity that the flow in the outer regions of the 
disk may also be non-axisymmetric.

For the sake of completeness, we point out that we apply this boundary condition
procedure directly to 
the mass densities. For the velocities, we take the additional steps of
converting the Cartesian velocity components back to the 
non-rotating frame, then constructing the radial and 
tangential components of the velocity from the Cartesian 
components, and performing the bilinear interpolation and
radial extrapolation procedure on these components. 
By performing the interpolation and extrapolation on the
radial and tangential components in the non-rotating frame, 
we first set the boundary conditions for quantities that are easier
to interpolate and extrapolate: namely the tangential velocity 
which is nearly flat (constant in x and y)  and the radial 
velocity which is nearly zero in the disk's outer regions 
in the nonrotating frame.

We stress that our boundaries
allow gas flow across them. For different runs, mass loss/gain 
after about 1 $\rm t_{\rm dyn}$ ranges from $\approx$ 2 per cent 
for runs performed in the non-corotating frame or slowly
corotating frames, to at most $\approx$ 15 per cent 
for the fastest corotating frame we simulated, i.e. when 
corotation is at $\approx$ 5 kpc.

\subsection{Steady State ?}\label{steadystate} 

Before proceeding to an examination of how different physical parameters
change the gas' response to the underlying gravitational
potential, we consider the question of whether
our conclusions depend on the specific snapshot in time for which we
analyze the simulation.
For this we look at (fig.~\ref{Longevol758})
the long term evolution of our ``fiducial'' model for NGC 4254.
Displayed are the density
contours at times 1, 2, and 3 Gyr. In terms of dynamical
time, this corresponds to 2.1 $\rm t_{\rm dyn}$,
4.2 $\rm t_{\rm dyn}$, and 6.4 $\rm t_{\rm dyn}$. 
We see that the density field adjusts to the force input within a couple of
dynamical times.
Therefore, our simulations
are even applicable to spiral arms which are not long-lived 
features, i.e. with ($t \approx t_{\rm dyn}$).
To illustrate the steadiness of
the features in time, we present a 
greyscale plot (lower
right hand corner of figure~\ref{Longevol758}) of the density
as a function of both time and azimuth along a 
circle of $\approx$ 3.8 kpc radius  (indicated by a
thick solid line in the contour plot at 3 Gyr).
The solid vertical line in the greyscale plot
indicates the time at which the full potential is turned on in the
simulation, i.e. $\rm t_{\rm FP} \approx$ 1.3 $\rm t_{\rm dyn}
\approx$ .62 Gyr.
Shortly after this moment (about 0.5 Gyr later), the morphology of the density
distribution becomes nearly time-independent. The contour plots show that 
even the orientation 
of the very inner region which we are not trying to model in detail
since it may have a different pattern
speed from the outer spiral pattern, seems to be steady in time.
What is true for the densities
applies also to the velocities. They also reach a near steady state
(cf. Fig. 6.3, Kranz 2002).

\subsection{Influence of Initial Density Profile}\label{initialdensity} 

To confirm that a specific choice of initial gas density profile
does not introduce a bias in our study, we perform simulations with 
initial density profiles
of double (7.72 kpc) and half (1.93 kpc) our fiducial initial 
scalelength of 3.86 kpc, as well as with an initial density 
profile which has a ``hole''
in the central region ($\rho_{\rm 0} (1 +\frac{r}{r_{\rm c}}) \exp(-\frac{r}{r_{\rm s}})$,
where $r_{\rm c} = \rm 4.46$ kpc and $r_{\rm s} = \rm 7.72$ kpc). 
Figure~\ref{morph_diffICS} and the top panel
of figure~\ref{diffICS_dens_rvel} reveal that while the density
contrast depends on the  initial density profile, 
the morphology of the final gas distribution is almost unaffected.
The bottom panel of 
figure~\ref{diffICS_dens_rvel} further shows that the density averaged 
radial velocity is also very nearly independent of the
initial density profile, except in a small region centered around the corotation 
radius (in this case $R_{\rm c} = 7.58$ kpc).
Note however that even in this region differences between models are slight.

   \begin{figure}
      \centerline{\psfig{file=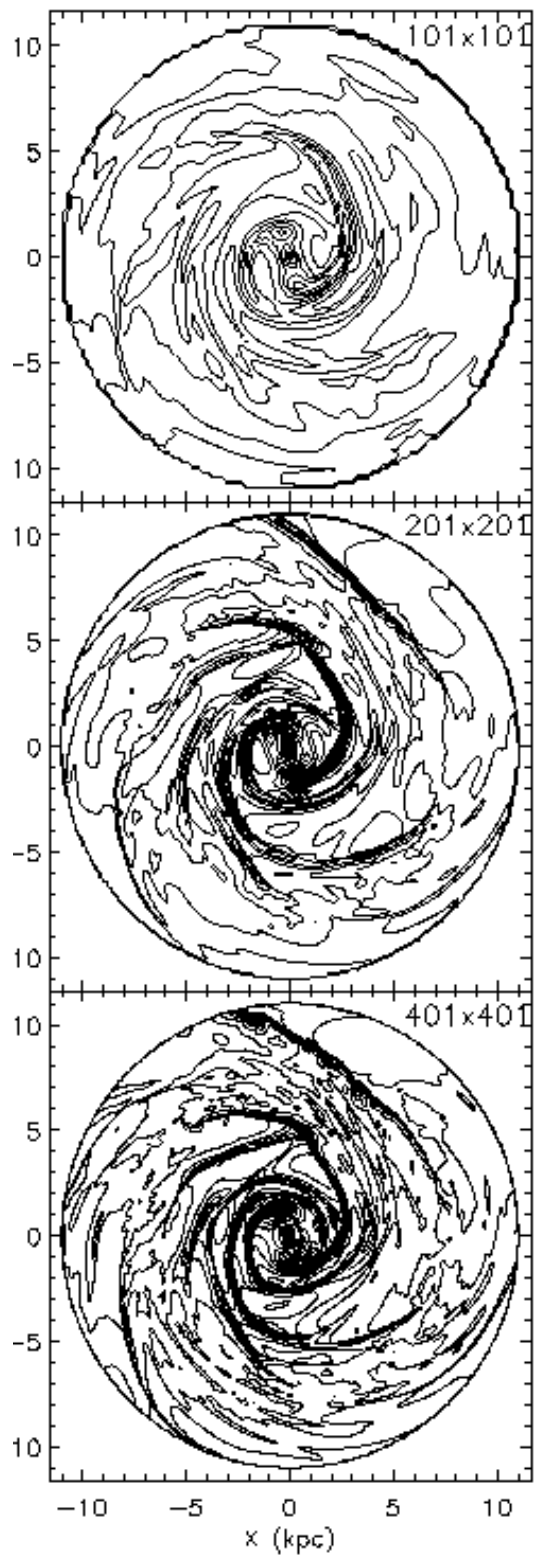,width=.85\hsize}}
      \caption{Contour maps of the density from
simulations with $c_{\rm s} = 10$ \kms, 60 per cent disk fraction,
$R_{\rm c} = 7.58$ $\rm{kpc}$, and with increasing
grid resolution. The full potential
is turned on in 20 cell sound crossing times for the 101 $\times$ 101
grid, 40 $t_{\rm s}$ for the 201 $\times$ 201 grid, and 80 $t_{\rm s}$ for the
401 $\times$ 401 grid. The result is shown after 1590 Myr ($\sim$ 3.3 $t_{\rm dyn}$) have elapsed.}
      \label{morph_res}
    \end{figure}

   \begin{figure*}
      \centerline{\psfig{file=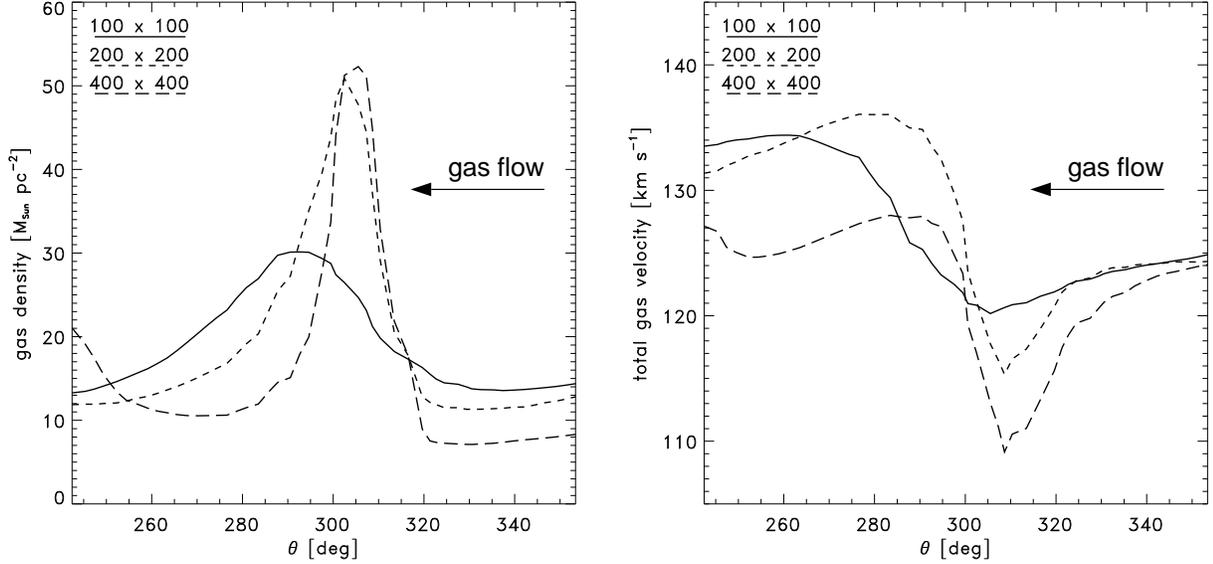,width=.9\hsize}}
      \caption{The gas density and amplitude of the velocity 
as a function of azimuth for $R \approx$ 3 kpc, for simulations performed
with 
$f_{\rm d} = 60$ per cent, $R_{\rm c} = 7.58$ $\rm{kpc}$, $c_{\rm s} = 10$\kms and different grid resolutions. }
      \label{shock_az_res}
    \end{figure*}
\subsection{Resolution} \label{resolution}

Results from any numerical study are also subject to 
the choice of grid resolution. Paper I was based on simulations
performed with an evenly spaced grid of 201 cells in x and in y
of length $\approx$ 115 pc per cell. Such a resolution quite closely
matched the observed kinematics obtained from the long-slit spectra.
Here, we run experiments using a grid with half and double this number of cells.
We find that results have fairly well converged for the 201 $\times$ 201 grid.
A plot of the density over the entire x-y plane, figure~\ref{morph_res},
for the sequence of increasing grid resolutions shows that the
simulation performed on the 101 $\times$ 101 grid is missing many
of the features present in the simulations on the 201 $\times$ 201 and the 401 $\times$ 401 grids.

More detailed examination of gas profiles in plots of the radial
density profile for $\vartheta = 202.5^\circ$ (figure~\ref{denscuts_all})
and of the azimuthal density profile for $r \approx$ 3 kpc (figure~\ref{shock_az_res})
reveals that the amplitudes of the density maxima for the 
201 $\times$ 201 grid have nearly converged to their values on the
401 $\times$ 401 grid.  There are however noteable differences between
the profiles on the different resolution grids. Firstly, the phases
of the density maxima are fairly well matched for the strong density maxima 
but less so for the lower density contrast ones.
For example, as shown near $R \approx 6$ kpc on figure~\ref{denscuts_all} 
higher resolution simulations tend to shift small density maxima to larger radii.
Secondly, as can be seen on figs.~\ref{shock_az_res} and~\ref{denscuts_all},
the shape of the density and velocity profiles changes with
resolution. The lowest resolution grid yields the smoothest and most 
symmetrical gas profiles. However with a grid resolution of
201 $\times$ 201 one already recognizes
the characteristic
profiles described analytically by Roberts (1969). The profiles are 
asymmetric with a rapid density rise followed by a gradual decline (see fig.~\ref{shock_az_res}).

Interestingly, even though the numerical viscosity 
increases with lower grid resolution, implying 
higher numerical gas inflow into the center, we
find that the density at $r=0$ for the 101 $\times$ 101 grid
(figure~\ref{denscuts_all})
is actually lower than the central densities of the higher
resolution simulations.  
We will discuss this result in more detail
in section~\ref{massinflow} but conclude here 
that the numerical errors associated with a 201 $\times$ 201 grid 
represent an $\approx$ 10 \% contribution to the shape, amplitude,
and position of the features of our discs.

\section{Parameter Study} \label{Parameters} 
Following these considerations about resolution, initial conditions
and the attainment of a quasi-steady state in the simulation, 
we focus on how the nature of gas flow in the potential is effected by
changes in the three parameters enumerated in section~\ref{hydro_modeling}:
the amplitude of the 
disk contribution to the potential $f_{\rm d}$, the gas temperature $c_{\rm s}$,
and the pattern speed $\Omega_{\rm p}$.
We explore only variations in each of these three parameters
individually, keeping the other ones
fixed at their fiducial values (given in boldface in Table 1).

\subsection{Different Disk Fractions} \label{diffdiskfractions}

At first we examine how increasing the non-axisymmetric stellar mass component 
of a gravitational potential influences the
simulated gas density distribution and the velocity field. The total
potential, $\Phi_{\rm{tot}}$, was assembled in the following way:
\begin{equation}
\Phi_{\rm{tot}}({\bf R}\mid f_{\rm d}) = f_{\rm d}\Phi_{\rm{stellar}}({\bf R}) +
\Phi_{\rm{halo}}({\bf R}\mid f_{\rm d})
\end{equation}
adopting the values 0.2, 0.45, 0.6, 0.85 and 1 for
$f_{\rm d}$. $\Phi_{\rm{stellar}}$ is the stellar potential with the maximal
stellar mass-to-light ratio and $\Phi_{\rm{halo}}$ is the potential of
the dark halo which is constrained by the observed rotation curve.
For a fair comparison, 
all the simulations in this series of increasing disk fraction
were run for the same amount of total time, namely $\sim$ 1.6 Gyr.

\begin{figure}
   \centerline{\psfig{file=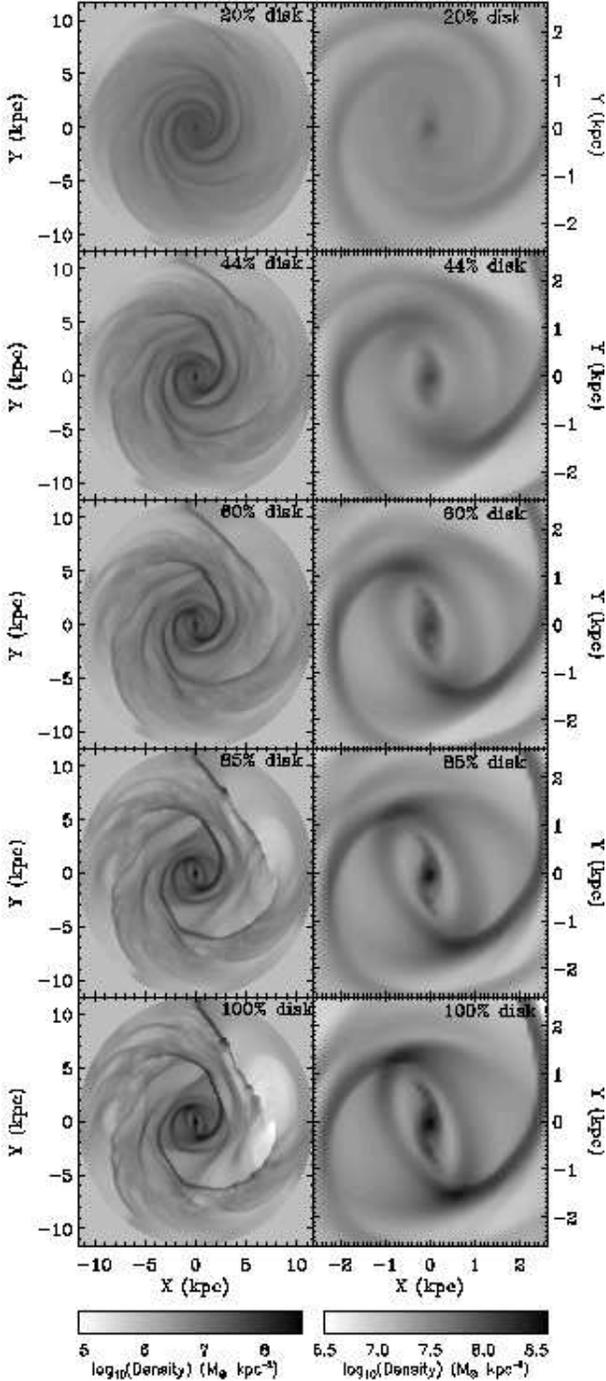,width=.95\hsize}}
      \caption{Grey scaled maps of the log of the gas density from
simulations on a 201 $\times$ 201 grid, with $c_{\rm s} =$ 10 km $\rm{s}^{-1}$, 
and $R_{\rm c} =$ 7.58 $\rm{kpc}$, and with increasing
disk contribution to the total galactic potential. The left column
shows the entire simulated region, and the right column shows only
the inner 2.6 kpc$^2$ region.}
      \label{morph_diskfrac}
    \end{figure}
   \begin{figure}
      \centerline{\psfig{file=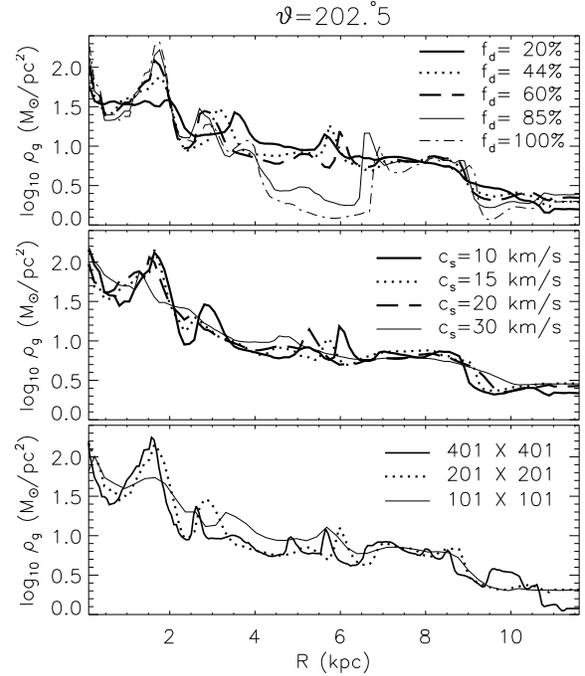,width=.9\hsize}}
      \caption{Cuts along an azimuth of $\vartheta = \rm 202.5^{\circ}$
      of the log of the density for runs with  $R_{\rm c} =$ 7.58 $\rm{kpc}$.
Top panel shows results from a simulation performed on a 201 $\times$ 201 grid
with $c_{\rm s} = 10 $\kms and varying disk fraction.
Middle panel shows results from a simulation on a 201 $\times$ 201 grid, with
a disk fraction of 60 per cent and varying sound speed. Lower panel shows
      results
from a run with 60 per cent disk fraction, a sound speed of 10 \kms, and
      varying
grid resolution. Each curve is displayed for the simulation at time $t=\rm{t_{\rm FP}} +2{\ }\rm{t_{\rm dyn}}$ ($\approx 1.6$ Gyrs).}
      \label{denscuts_all}
    \end{figure}

A logarithmic plot (figure~\ref{morph_diskfrac}) of the gas density 
for the sequence of simulations,
shows that the density contrast of the non-axisymmetric features in
the gas increases as the disk fraction increases.
This corroborates Figure~\ref{wiggle_amplitude} in this paper
and Figure 8 of Paper I 
which quantified this trend 
by taking the average of the
amplitude of the velocity deviations from axisymmetry, and found
that this average increases more or less linearly with the
disk fraction. In addition to the change in the density contrast, 
figure~\ref{morph_diskfrac} shows that
features become more ``angular'' with increasing disk fraction.
For example, in the lowest disk fraction case (20 per cent disk), 
the spiral arms appear to be rounded and
smooth in their curvature. As the disk fraction
is increased to 85 per cent,  the lower spiral arm develops a squareness
which is even more pronounced for the 100 per cent disk fraction case.

\begin{figure*}
      \centerline{\psfig{file=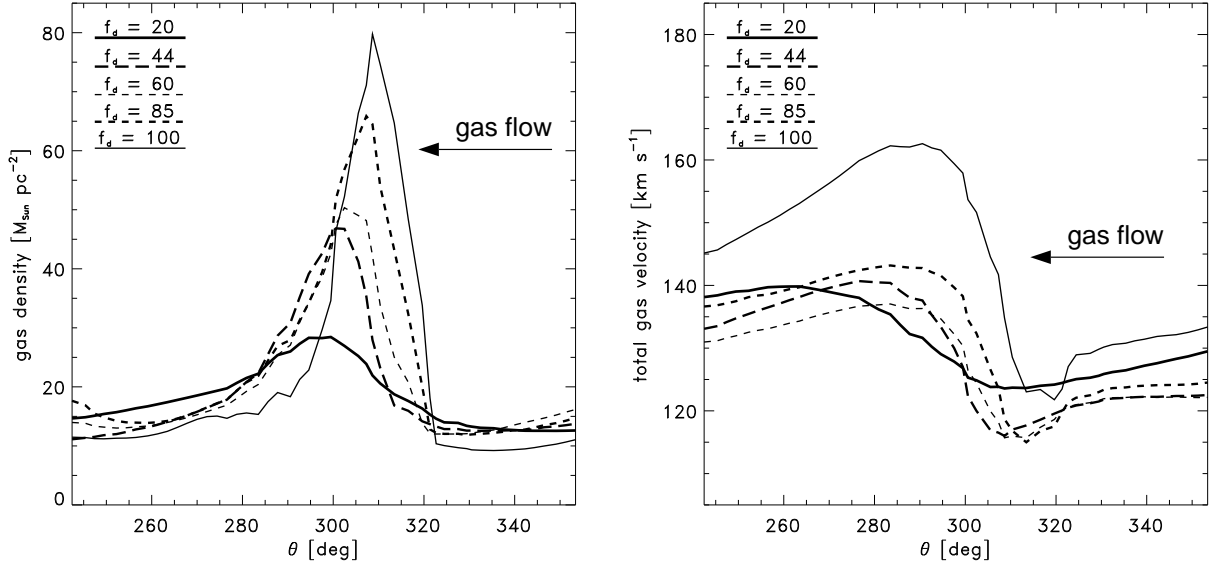,width=.9\hsize}}
      \caption{The gas density and amplitude of the velocity 
as a function of azimuth for $R \approx $3 kpc, for simulations performed
on a 201 $\times$ 201 grid with 
$c_{\rm s} = 10 $\kms, $R_{\rm c} = 7.58 \rm{kpc}$ and different values for $f_{\rm d}$. }
      \label{shock_az_fd}
    \end{figure*}

To display these changes in amplitude and morphology in more detail,
figure~\ref{denscuts_all} plots the density profile along 
an azimuthal cut through the disk at a position angle \footnote{Figure
1 of Paper I displays the orientation of the cuts along all discussed
position angles.} $\vartheta = \rm 202.5^{\circ}$.
The top panel of this figure shows that the smaller amplitude 
density peak (lower spiral arm) located at
$R \approx$ 6 kpc
moves outwards with increasing disk fraction 
suffering larger shifts in position with changing
disk fraction than the larger amplitude peaks at smaller radii
($R < $ 2 kpc). To further quantify this, we plot the gas density and velocity 
amplitude as a function of azimuth for $R \approx$ 3 kpc (figure~\ref{shock_az_fd}). 
In agreement with Woodward's (1975) analytic calculations (his figure 7), 
we find that the location of the density peak moves towards larger azimuths
with increasing perturbation strength. 
Figure~\ref{denscuts_all} also indicates that
as the disk fraction increases, regions of
increasingly lower gas density appear immediately adjacent to the gas 
density peaks in the arms.
Hence to differentiate between models with different $f_{\rm d}$
a code has to perform well in the low
density regions, which is one of the strengths of
grid codes over particle codes in general and of the BGK scheme
in particular.

   \begin{figure*}
      \centerline{\psfig{file=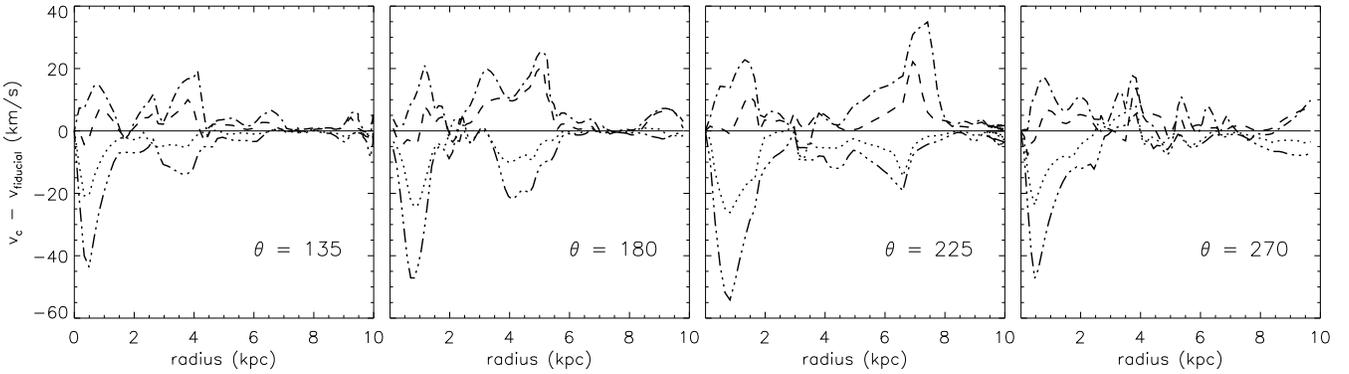,width=.32\hsize,angle=270}}
      \caption{Cuts along different position angles of the difference
between the velocity amplitude for runs with different disk
fractions compared to a 'fiducial' run with a 60 per cent disk fraction.
All simulations referred to in this plot are done on 
a 201 $\times$ 201 grid, with a sound speed of 10 km $\rm s^{-1}$ and 
$R_{\rm c} = 7.58$ kpc. 
The dash-dot-dot-dot line is for the 20 per cent disk, the dotted line
is for the 44 per cent disk, the dashed line is for the 85 per cent disk
and the dash dotted line is for the 100 per cent disk. }
      \label{velcuts_diskfrac}
    \end{figure*}

Morphological changes with increasing
disk fraction are not limited to spiral arms: they are also present in 
the inner region of the galaxy.
Even at the resolution of our study, 
the grey-scale maps of the inner regions of the 
disk (right column of figure~\ref{morph_diskfrac}), 
reveal the development 
of off-axis shocks, the emergence of an oval ring around the inner
region, and the strengthening of the 4/1 shock at the endpoints
of this oval ring.

Since we are
ultimately comparing the kinematical information in the simulations
to observations in order to constrain the dark matter fraction of the galaxy, 
we plot in figure~\ref{velcuts_diskfrac}
the difference between velocity amplitudes
($\sqrt{{v_x}^{2}+{v_y}^{2}}$) 
measured in each model  and those measured in 
our fiducial run ($\rm f_{\rm d} = 60$ per cent)
along four position angles ($\vartheta = \rm
135^{\circ},\rm 180^{\circ}, \rm 225^{\circ}, \rm 270^{\circ})$. 
As already seen in figures~\ref{morph_diskfrac}, ~\ref{denscuts_all} as well
as in figure 7 of paper I, there are significant variations 
in the models depending on the assumed value for $\rm f_{\rm d}$. 
The additional information contained in figure~\ref{velcuts_diskfrac} is
that changing $\rm f_{\rm d}$ has a greater impact on the central
region of the disk where the differences in velocities between
different models can reach 70 \kms.
However there are also large velocity variations (10 - 40 \kms)
throughout the rest of the disk.

Another feature which reacts to the changing disk fraction is
the shock touching the boundary in the upper right hand quadrant. 
It becomes more inclined towards the center of the disk
and increases in length with increasing $\rm f_{\rm d}$, raising
the concern that it might interact with other features
in the disk. It is hard to discount this shock as artificial 
because Figure~\ref{Potential} indicates that
there is a minimum in the potential in the upper quadrants of the
grid. However it is likely that whatever structure forms in that area
of the grid, may be affected by the outer boundary conditions.

\subsection{Different Sound Speeds} \label{diffsoundspeeds}

   \begin{figure}
      \centerline{\psfig{file=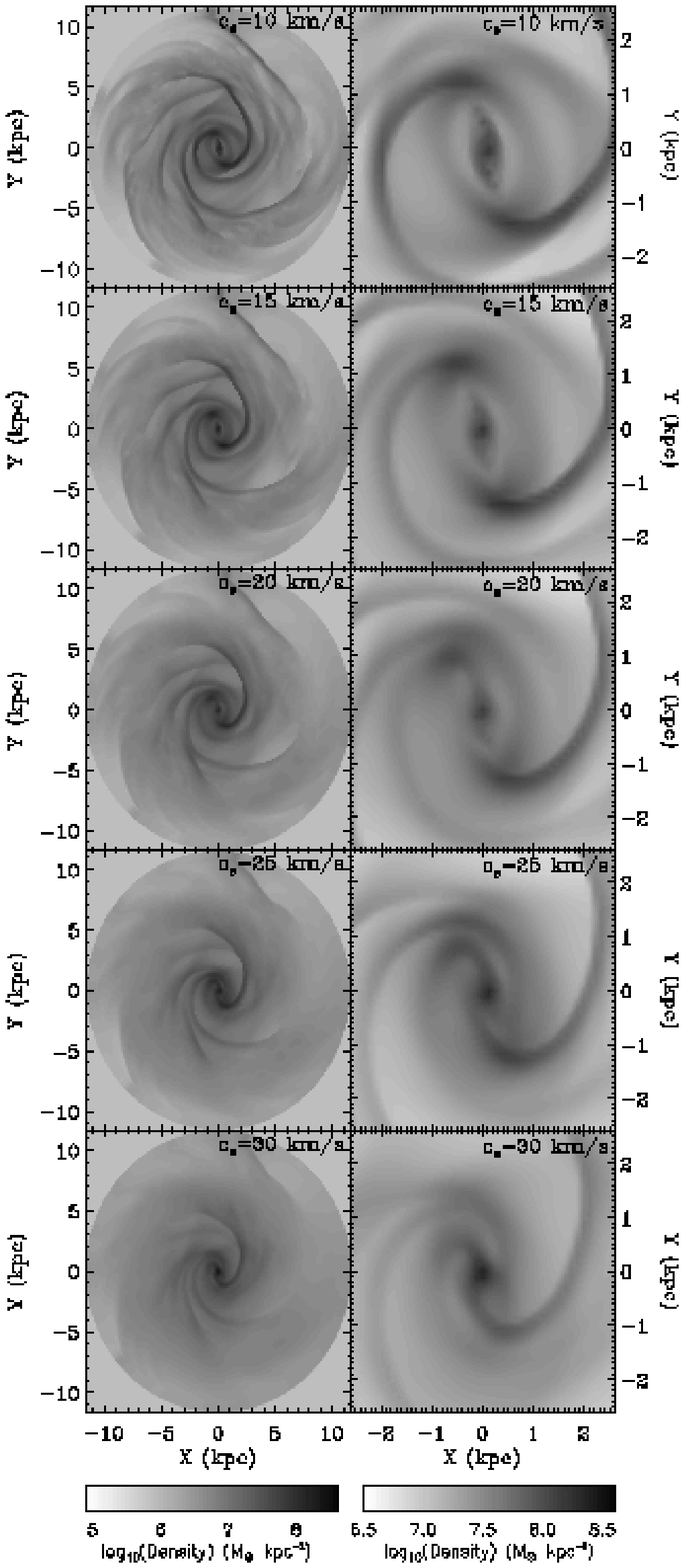,width=.95\hsize}}
      \caption{Grey scaled maps of the log of the density from
simulations on a 201 $\times$ 201 grid, with 60 per cent disk fraction,
$R_{\rm c} =$ 7.58 kpc, and with increasing
sound speed. The entire simulated region is shown in the left column
      and the inner 2.6 kpc$^2$ region is shown in the right column.}
      \label{morph_cs_full}
    \end{figure}

   \begin{figure*}
      \centerline{\psfig{file=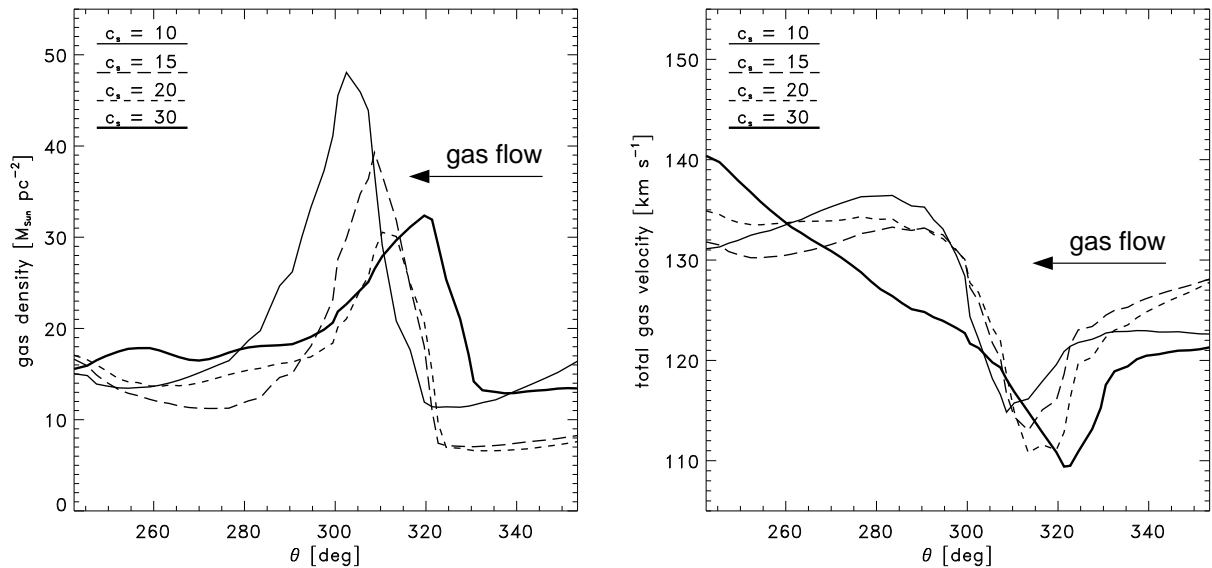,width=.9\hsize}}
      \caption{The gas density and amplitude of the velocity 
as a function of azimuth for $R \approx $3 kpc, for simulations performed
on a 201 $\times$ 201 grid with 
$f_{\rm d} = 60$ per cent, $R_{\rm c} =$ 7.58 kpc and different values for $c_{\rm s}$. }
      \label{shock_az_cs}
    \end{figure*}

   \begin{figure*}
      \centerline{\psfig{file=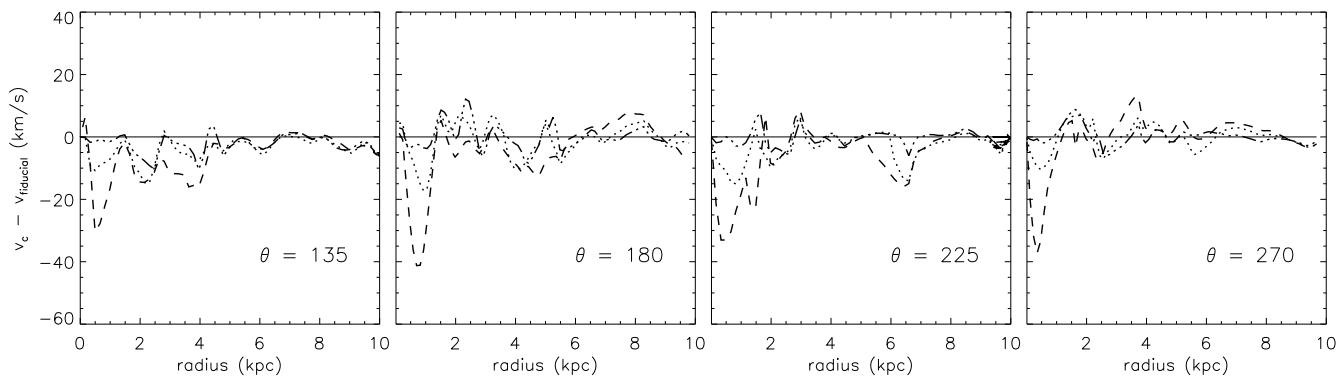,width=.32\hsize,angle=270}}
      \caption{Cuts along different position angles of the difference
between the velocity amplitude for runs with different sound
speeds compared to a 'fiducial' run with a sound speed of 10 km s$^{-1}$.
All simulations referred to in this plot are done on 
a 201 $\times$ 201 grid, with a disk fraction of 60 per cent and 
$R_{\rm c} = $7.58 kpc. 
The dash-dot-dot-dot line is for $c_{\rm s} = 15 $\kms, the dotted line
is for $c_{\rm s} = 20$\kms, and the dashed line is for 
$c_{\rm s} = 30 $\kms.}
      \label{velcuts_cs}
    \end{figure*}

Because it is thought that the interstellar cloud medium can be 
crudely approximated by an isothermal gas if the clouds have an 
equilibrium mass spectrum (Cowie 1980) maintained by supernovae
which both destroy and create gas clouds, most simulations of gas flows
in disk galaxies treat the gas as isothermal where the sound speed
represents the rms velocity of the interstellar clouds.  
Simulations treating the gas as multi-phase are starting to be run 
(Colina \& Wada 2000, Wada \& Koda 2001, Slyz et al. 2003). Given 
the caveats in modeling a multi-phase interstellar medium,
we find it prudent to keep the simple assumption of
a uniform ISM for these global disk simulations,
since, as already remarked in the Introduction, for our 
study in Paper I we are trying to match velocity wiggles in the observations 
to those in simulations
in as much detail as possible. As a matter of fact it is virtually impossible
to model star formation and feedback processes
in our simulations in such a way as to
match the velocity wiggles in the velocity spectra.
In other words, our working hypothesis
is that these wiggles arise from variations in the gravitational potential and
therefore that an isothermal equation of
state is a good description of the ISM.
We are then left with examining the 
effect on our results of different assumptions for
the uniform sound speed of the gas.

Typically authors assert that 
their hydrodynamical calculations are insensitive to reasonable
changes in the sound speed
(Lindblad, Lindblad, \& Athanassoula 1996,
Lindblad \& Kristen 1996, Weiner et al. 2001). Given that
flow velocities of the gas relative to
the bar and/or spiral pattern greatly exceed the velocity dispersion
of the interstellar medium throughout almost the entire disk of a
galaxy, it indeed seems reasonable to think that there should be no strong
dependence of the flow on the sound speed.
At larger values of the sound speed ($\approx$25 km s$^{-1}$) however,
detailed investigations 
of gas flows in strongly barred galaxies (Englmaier \& Gerhard 1997;
Patsis \& Athanassoula 2000), show that the structure
of the flow changes markedly.

The first effect of increasing the sound speed that one expects
to find is that the gas should respond less strongly to the
forcing pattern, since at higher sound speeds the pressure of
the gas starts to become more important.
This effect is evident in figure~\ref{morph_cs_full} where in 
a sequence of simulations
differing only in their sound speed we see the non-axisymmetric
features in the gas gradually fade with increasing sound speed.
By $c_{\rm s} = 30$ km s$^{-1}$ the spiral structure does not extend
as far as it does in the colder gas runs even though
traces of some of the larger spiral features are still present.
We emphasize that sound speeds of 25 or 30 km s$^{-1}$ throughout
the entire disk are unrealistically high. Results from simulations
at these high values are merely included to illustrate trends of
increasing sound speeds.

A close-up view (right hand column of figure~\ref{morph_cs_full}) 
of the interior 
region of the simulation shows even more striking morphological
differences between simulations at different sound speeds.
Firstly we notice that the prominent spiral arm on the right
side of the galaxy winds up more tightly with increasing sound speed.
Secondly we see that the spiral arms reach further
and further into the center of the disk until by 
$c_{\rm s} = $ 30 km s$^{-1}$ they
are completely connected to the centermost region.
The last principle morphological change we see is that the
4/1 shocks fade with increasing sound speed until they vanish
by 25 km s$^{-1}$.
Although the gravitational potential we are working with is
different from the one used by Englmaier \& Gerhard (1997) and 
Patsis \& Athanassoula (2000), some of these changes that
we see, namely the disappearance of the off-axis shocks and the
fading of the 4/1 shocks, are similar to those they describe.

For a more detailed view of the influence of changing sound
speed on the gas flow, we display the density profiles for the 
different runs at $\vartheta = \rm 202.5^{\circ}$ (figure~\ref{denscuts_all}).
As expected the density profile for the simulation at $c_{\rm s} = $ 30 km s$^{-1}$ is 
everywhere much smoother than the density profiles for
the other simulations. The density maximum in
the middle of the disk ($R \approx 4 -5$ kpc) is present for
all the runs but it progressively shifts radially inwards 
with increasing sound speed which is a manifestation of the stronger
winding of the spiral structure for larger sound speeds that was
already mentioned above.
Features between 2 and 3 kpc
which are present for $c_{\rm s} = 10$  km s$^{-1}$ are essentially gone by 
$c_{\rm s} = $25-30 km s$^{-1}$. Another effect seen in this figure, 
is that at $r = 0$, the simulations with sound speeds between
10-20 km s$^{-1}$ achieve effectively the same density, but
the density for the run with $c_{\rm s} = 30$ km s$^{-1}$ is higher
by a factor of $\approx 1.09$ in log-space.In a similar way as in 
figure~\ref{shock_az_fd} for the disk fractions, we plot the gas density
and amplitude of the velocity for changing sound speed in
figure~\ref{shock_az_cs}. The density peak 
is systematically shifted towards larger
azimuths for increasing sounds speeds again reflecting a tighter winding
of the spiral, with a shift of $\approx 20^{\circ}$
between the smallest (10 km s$^{-1}$) and largest (30 km s$^{-1}$) sound speed 
we considered which is larger than the $15^{\circ}$ shift induced by changes
in $f_{\rm d}$. However changing
the disk fraction from 20 per cent to 100 per cent results in a difference of 53 
$M_\odot$ $\rm{pc}^{-2}$ in density amplitude, as compared to a difference of 16 $M_\odot$ $\rm{pc}^{-2}$ for
a variation in sound speed between 10 and 30 km s$^{-1}$. Thus
the simulations are less sensitive in this respect to a change in $c_{\rm s}$ as compared to
changes in $f_{\rm d}$.

Figure~\ref{velcuts_cs} shows the difference between 
velocity amplitudes for runs with different sound speeds compared to the run with
$c_{\rm s} = 10$ \kms, for the same four position angles used in 
figure~\ref{velcuts_diskfrac}. As was the case for the different $f_{\rm d}$
models, changing the sound speed 
has the greatest impact on the centermost region of
the galaxy with velocity variations of up to $\approx$ 40 km s$^{-1}$.
This is worrisome because it is similar to the variation in this
region between the models for different disk fractions .
However, if one excludes the centermost region of the disk, then
variation in the velocity amplitudes as one changes 
the sound speed is much lower than the variations between the models
for different disk fractions (at most $\approx 15$ km s$^{-1}$ as compared to 
$\approx 40$ km s$^{-1}$)
This lower sensitivity to changes in sound speed
in the face of uncertainties in how to model the interstellar medium, 
suggests that the gas response in the outer regions, as opposed to the
inner regions, of the disk may more reliably trace the gravitational
potential. Furthermore since a sound speed of 10 km s$^{-1}$ for the
interstellar medium is physically motivated, we maintain that
for simplicity's sake, it is a good choice for these kind of studies.

   \begin{figure*}
      \centerline{\psfig{file=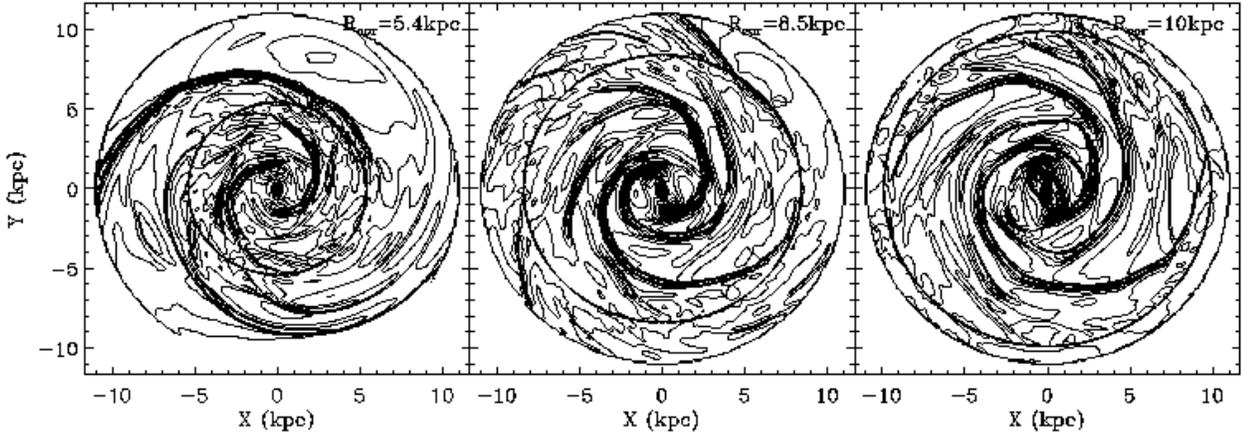,width=\hsize,angle=0}}
      \caption{Density contours for  3
simulations differing in their pattern speed. The value for the
      corotation radius is marked in the upper right hand corner.
All the simulations are performed
for a model with 60 per cent disk fraction, $c_{\rm s} = 10$ km s$^{-1}$ and on a
201 $\times$ 201 grid.}
      \label{effpotvel}
    \end{figure*}
   \begin{figure*}
      \centerline{\psfig{file=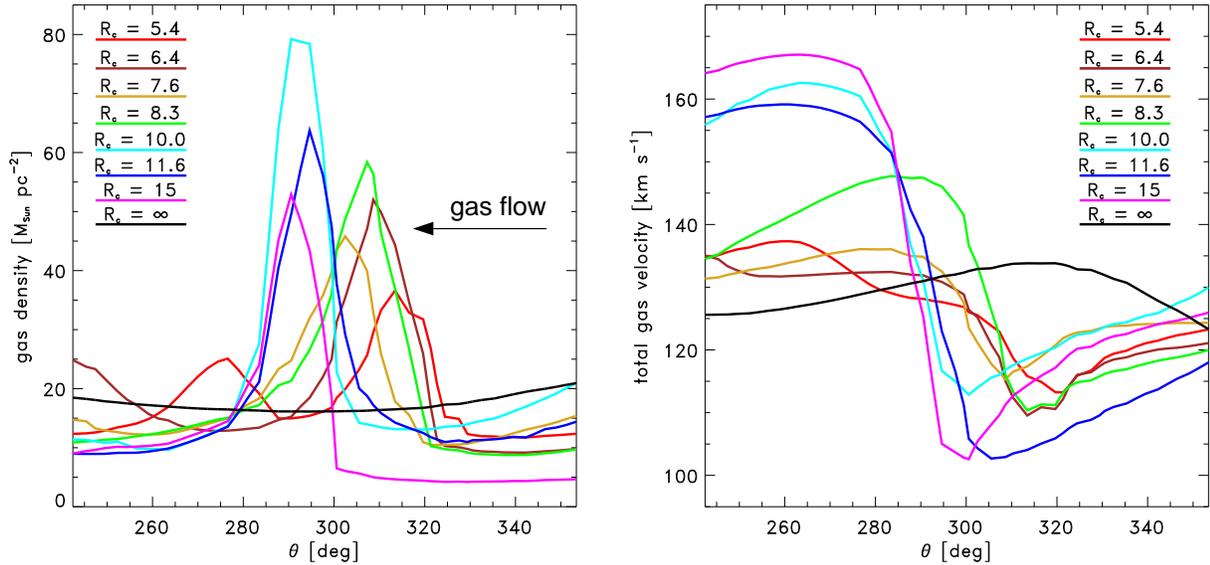,width=.9\hsize}}
      \caption{The gas density and amplitude of the velocity 
as a function of azimuth for $R \approx $3 kpc, for simulations performed
on a 201 $\times$ 201 grid with 
$c_{\rm s} = 10 $\kms, $f_{\rm d} = 60$ per cent, and a range of values for $R_{\rm c}$. }
      \label{shock_az_rc}
    \end{figure*}

\subsection{Different Pattern Speeds} \label{patternspeeds}

Bar simulations have already shown that the most significant parameter
in controlling the structure of the gas flow in a disk is its angular pattern
speed, $\Omega_{\rm p}$ (e.g. Hunter et al. 1988). Given this sensitivity
of the gas morphology to the pattern speed (see figures~\ref{effpotvel} 
and~\ref{shock_az_rc}), matching the simulated
density to an observation of the galaxy surface density may be a
powerful way to constrain it (see e.g. 
Garcia-Burillo et al. 1993, Garcia-Burillo et al. 1994, Sempere et al. 1995,
Mulder \& Combes 1996, Paper I). 

To elaborate, simulations in a fixed gravitational potential 
show that one of the things that the pattern speed and hence the corotation
radius determines is the radial extent of the spiral pattern in the gas.
This is easy to understand because the corotation radius is the radius 
at which the pattern speed $\Omega_{\rm p}$ is equal to the 
orbital frequency $\Omega$. Hence at the corotation resonance (see 
figure~\ref{resonances}) the gas rotates along with the spiral 
perturbations and therefore the non-axisymmetric forcing vanishes. 
This causes spiral disturbances to be highly damped at corotation 
(see figure~\ref{effpotvel} for three cases of $\Omega_{\rm p}$).
Outside corotation the spiral structure may resume.
In light of this, it is expected (Shu, Milione \& Roberts 1973) 
that at the corotation radius star formation cannot get excited
by the density wave and should not be observed in a quiescent galaxy. 
Our best matching simulation for NGC 4254 (see Paper I) is consistent
with this signature for corotation. It gave a 
corotation radius beyond which star formation, i.e.~the
occurrence of H II regions, was largely reduced.

Nevertheless, a major caveat of both this approach for determining 
the pattern speed, and of our assumption of a single pattern
speed when we perform simulations to determine the disk fraction
of spirals is that spiral galaxies have a unique
pattern speed which is constant in time.
If spiral patterns are transient as indicated by some N-body simulations
(e.g.~Sellwood \& Carlberg 1984; Sellwood 2000) then the concept of 
established resonances is less
important, and the corotation radius might change very rapidly. Since
we model the gas flow in a fixed potential, we cannot explore the time
evolution of the spiral patterns. However in view of
the well ordered spiral structure of NGC 4254, and the fact
that the simulated gas density distribution accurately matches the 
observed galaxy morphology, the assumption that the spiral pattern is not
undergoing a massive rapid reorganization 
seems reasonable. We therefore conclude
that a stellar spiral pattern rotating at a unique speed $\Omega_{\rm p}$
is a sensible assumption for our simulations.

   \begin{figure}
      \centerline{\psfig{file=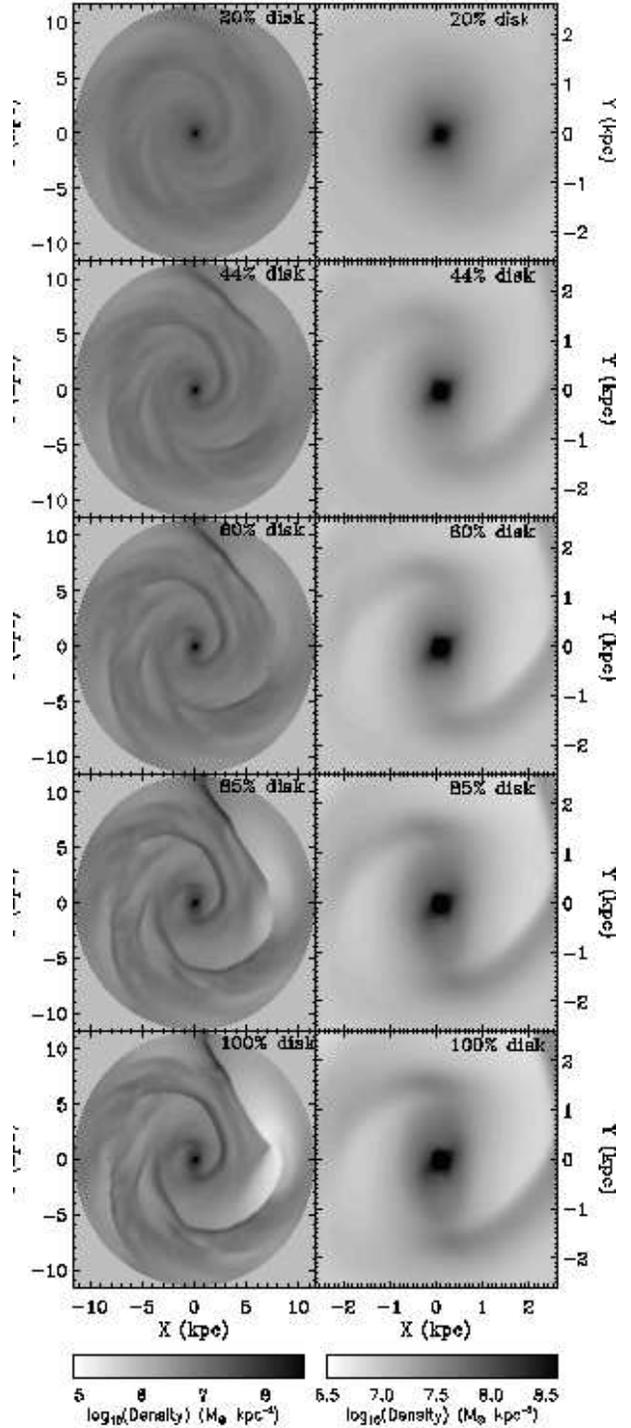,width=.95\hsize}}
      \caption{Grey scaled maps of the log of the density from
simulations with the beam scheme on a 201 $\times$ 201 grid, with 60 per cent disk fraction,
$R_{\rm c} = $7.58 kpc, $c_{\rm s} = 10$\kms and with different disk fractions. The entire simulated region is shown in the left column
      and the inner 2.6 kpc$^2$ region is shown in the right column.}
      \label{morph_beam}
    \end{figure}

   \begin{figure*}
      \centerline{\psfig{file=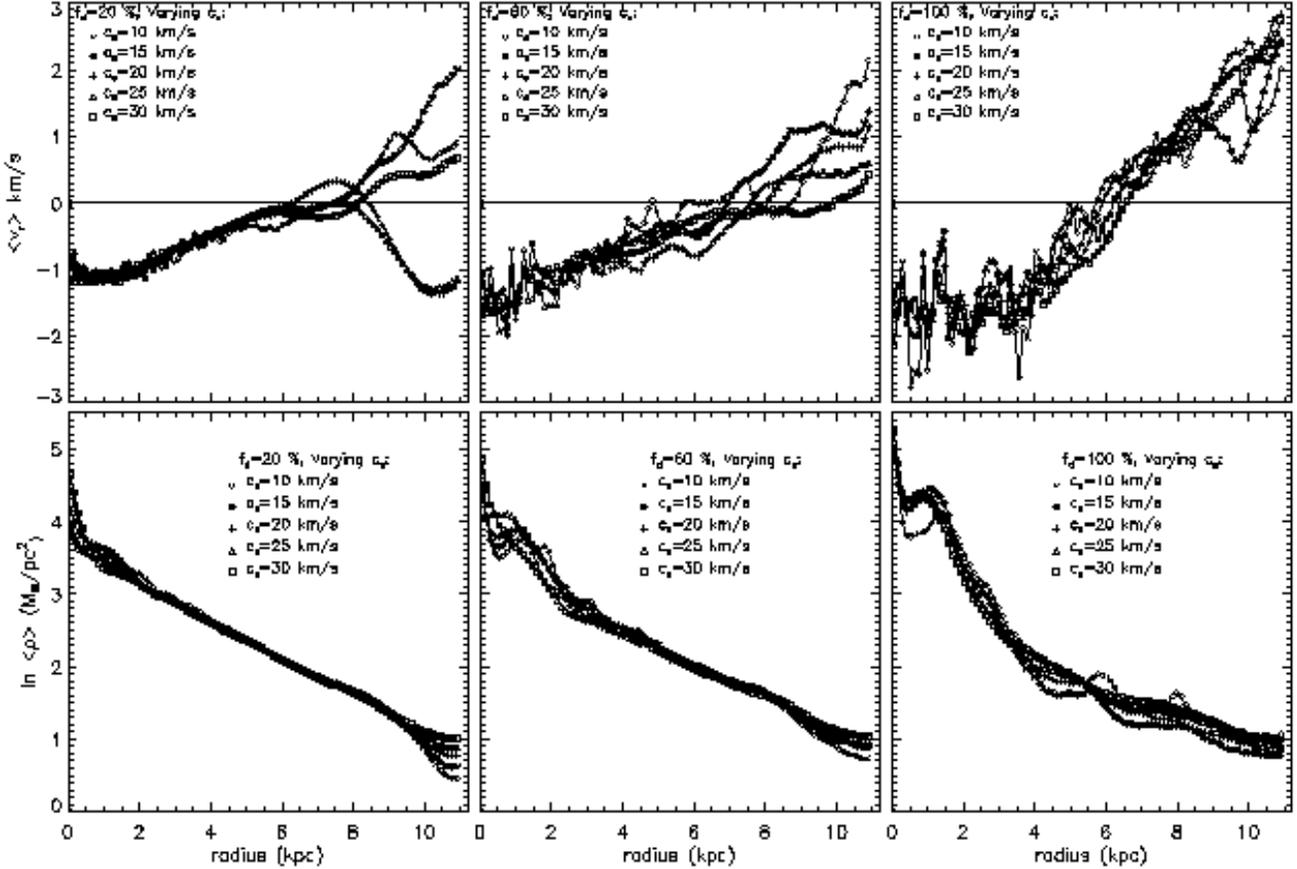,width=\hsize}}
      \caption{Plots of the density averaged radial velocity (top row)
and corresponding natural logarithm of the average density (bottom row) for
simulations on a 201 $\times$ 201 grid, with 20, 60 and 100 per cent disk fraction,
$R_{\rm c} = $7.58 kpc, and $c_{\rm s}$ ranging from  $10$\kms to $30$\kms.}
      \label{diffcs_difffd}
    \end{figure*}

   \begin{figure}
      \centerline{\psfig{file=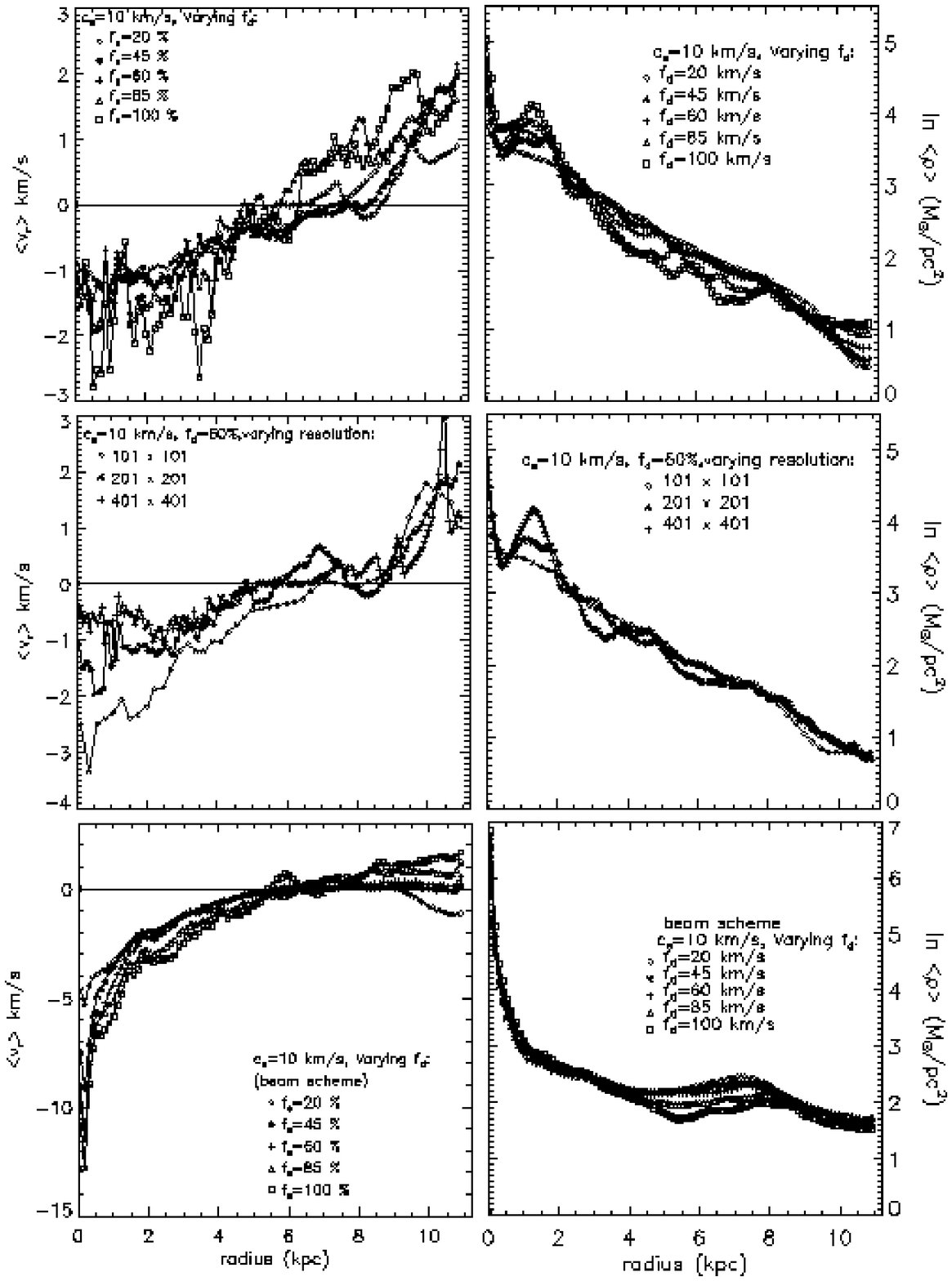,width=1.2\hsize}}
      \caption{Plots of the mass--averaged radial velocity (left column)
and corresponding natural logarithm of the average density (right column) for
simulations on a 201 $\times$ 201 grid with $R_{\rm c} =$ 7.58 kpc and $c_{\rm s}=10$\kms.
Top and bottom rows show results from simulations  with disk
fraction ranging from 20 per cent to 100 per cent performed with the BGK 
and beam scheme respectively. Middle row shows results from the BGK scheme for
the 60 per cent disk case, with different grid resolutions.}
      \label{difffd_codes_res}
    \end{figure}

\section{Mass Inflow} \label{massinflow}

Another way to quantify the influence of the different simulation
parameters is to consider the mass inflow rates.
As pointed out by Athanassoula (1992), a good indicator of the mean
inflow rate is the mass-averaged radial velocity, $<v_r>$.
Hence we compute it (fig.~\ref{diffcs_difffd}, top row;
fig.~\ref{difffd_codes_res}, left column) as a function of radius
for different simulations. We also plot the
average densities, $<\rho>$, (fig.~\ref{diffcs_difffd}, bottom row;
fig.~\ref{difffd_codes_res}, right column) so as to illustrate the
net effect of the radial inflow velocity on the gas mass distribution.

Fig.~\ref{diffcs_difffd} shows that changing the gas sound
speed affects the simulations with a high disk fraction more than
the simulations with a low disk fraction. More specifically,
the scatter about the average radial velocity is $\approx$
1 \kms for the 100 per cent disk case as opposed to
$\approx$ 0.1 \kms for the 20 per cent disk case. 
The figure also shows that the maximum of $<v_r>$
varies from $\approx$ --1 \kms for the 20 per cent disk case 
to $\approx$ --2 \kms for the 100 per cent disk case. 
The larger $|<v_r>|$ for the higher
disk fraction cases can be explained by the fact that 
increasing the disk fraction increases the
strength of the spiral shocks. This in turn increases the amount
of dissipation and hence angular momentum loss that the gas suffers 
as it slams into the shocks during its rotation about the galactic center

Fig.~\ref{diffcs_difffd} also shows that the shape of
the $<v_r>$ profile changes with
increasing disk fraction. More specifically 
for the 100 per cent disk simulations the region of the disk where
$|<v_r>|$ is maximum extends from the center to 
$\approx$ 3.5 kpc. As a consequence of this high inflow up 
to large distances
the average scale length of the
density distribution for the region interior to $\approx$ 3.5 kpc 
is much smaller than the initial gas radial scale length. In
contrast, for the 20 per cent disk fraction case only the 
region contained within the inner 2 kpc of
the disk departs in shape from the initial condition.

In addition to sensitivity to disk fraction and sound speed, 
the mass inflow rates, as pointed out by Prendergast (1983), are 
incredibly sensitive to code and grid spacing.
To explore this issue, we measured $<v_r>$ and $<\rho>$ for
simulations with different grid resolutions and also for a set of
simulations done with a different code.
As a worst case scenario for a diffusive grid code we took the 
beam scheme (Sanders and Prendergast 1974) which was used
extensively in the earliest galactic disk simulations (Huntley 1978,
Sanders \& Tubbs 1980, Duval \& Athanassoula 1983 ). Like BGK,
the beam scheme is a gas-kinetic hydrocode, i.e. fluxes are computed
by taking moments of a velocity distribution function, $f$. 
Both schemes arbitrarily choose $f$ at the beginning of
each updating time-step but the beam scheme evolves it through the
collisionless Boltzmann equation whereas the BGK scheme solves for 
the time evolution of $f$ throughout an updating time-step using the BGK
equation which is a model of the collisional
Boltzmann equation. 
By assuming instantaneous relaxation of $f$ to a Maxwellian velocity
distribution at the beginning of the updating time-step, the beam scheme
endows the gas with a mean collision time equivalent to the updating
time-step. In the BGK scheme, on the other hand, collisions are
active throughout the updating time-step and for hydrodynamical
applications the BGK scheme demands that the collision time be
much smaller than the updating time-step. Since dissipation
parameters are proportional to the collision time, $\tau$, e.g. the
dynamical viscosity $\eta = \tau$ p , we easily see that an
overestimation of the collision time will lead to a very diffusive
scheme. Indeed, fig.~\ref{difffd_codes_res} illustrates that $<v_r>$
is about an order of magnitude larger with
the beam scheme than it is with BGK and that the effect is worse in 
the center of the disk where a plot of $<\rho>$ shows that simulations
with different disk fractions are indistinguishable. Figure~\ref{difffd_codes_res} also shows that even with a BGK 
simulation on a 101 $\times$ 101 grid one still does better than beam 
by about a factor of 5 in the radial inflow velocities in the inner regions 
of the disk.

A look at the morphology of the disk for simulations performed with the beam scheme, (figure~\ref{morph_beam}) confirms that the inner
region of the disk computed with beam is essentially insensitive 
to disk fraction. The
morphological changes that accompany a change in disk
fraction (Fig.~\ref{morph_diskfrac}) are absent. Interestingly
however, figures~\ref{difffd_codes_res} and~\ref{morph_beam} show 
that even with this incredibly diffusive
scheme, one can still distinguish between simulations run with
different disk fractions in the outer regions of the disk where 
a spiral perturbation is present.

\section{Discussion} \label{discussion}

\begin{figure}
 \centerline{\psfig{file=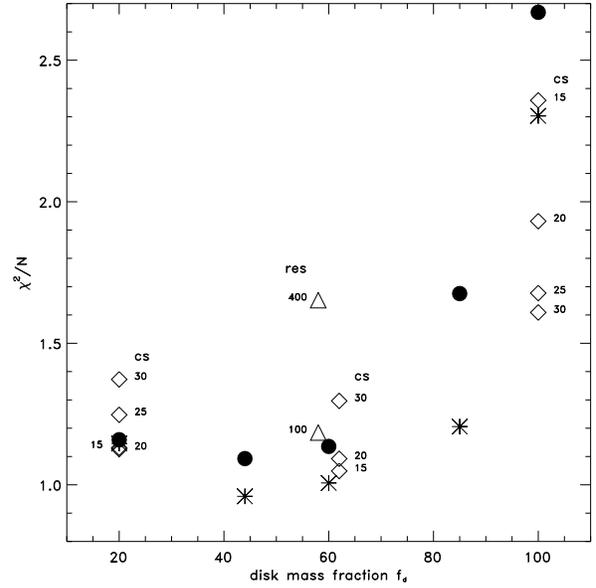,width=\hsize}}
 \caption{$\chi^2$ comparison between simulations and observations for
the set of simulations explored in this paper. All points, with the
 exception of the large solid points (beam scheme), are from
 simulations with the BGK scheme. The asterices denote simulations
 performed on a 201 $\times$ 201 grid, with $c_{\rm s} = 10$\kms and $R_{\rm c} = 7.58$
 kpc. Open triangles indicate simulations at different resolution, and
open diamonds indicate $\chi^2$ results from simulations at different sound speeds.}
 \label{chi}
\end{figure}
\begin{figure}
 \centerline{\psfig{file=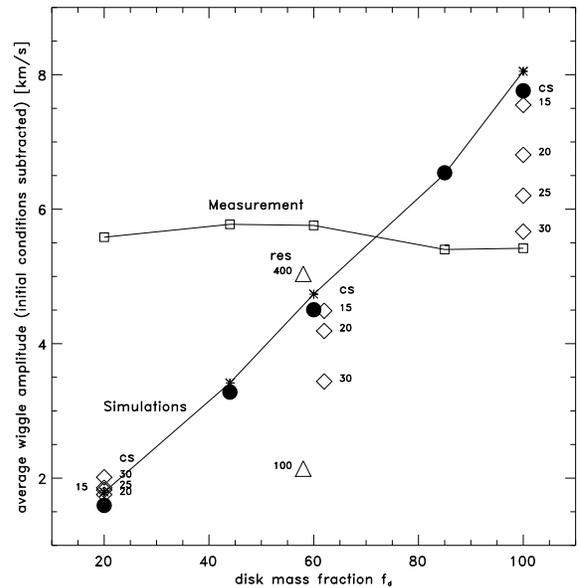,width=\hsize}}
 \caption{Plot of the average deviation of the velocity from
 axisymmetry for both the observations (open squares) and the simulations
 (asterices, filled circles, open triangles, open diamonds). The symbols for 
the simulations have the same meaning as the symbols
 in figure~\ref{chi}.  An axisymmetric model was fit to the observation, and
 then this model fit was subtracted from both the observation, and the
 simulation. Due to differences in the axisymmetric model fit for the
different disk fractions, there is a small variation in the average
 wiggle amplitude for the measurements. }
 \label{wiggle_amplitude}
\end{figure}
In order to quantify what has been described in Sections~\ref{Parameters} 
and~\ref{massinflow} we perform an overall comparison between the 
observed and simulated kinematics. We use two approaches. First we 
perform a global least squares
analysis (figure~\ref{chi}) taking into account the location \emph{and} 
amplitudes of individual wiggles. Secondly, we compute the average wiggle 
amplitude (figure~\ref{wiggle_amplitude}), neglecting information 
about the wiggle
positions. The comparison was performed on a reduced data set with 
the very inner disk region removed. Furthermore, a treatment has 
been applied to the 
observed kinematics in order to exclude shocks that are missing 
from \emph{all} simulations. These 
shocks probably originate from non-gravitational effects
(see Kranz 2002, \S 4.3.2.1). 

Figure~\ref{chi} corresponds to figure~9 in Paper I (except that 
it uses the reduced data set described above) 
and measures the match in both wiggle position and amplitude
between the observations and simulations.
Our BGK simulations with fiducial values for the grid resolution, 
gas sound speed, and pattern speed are plotted with asterices.
As argued in Paper I, a maximal disk mass fraction can be ruled out for 
NGC 4254 on the basis of this plot.
Now we can look at how strongly the locations of the points in figure~\ref{chi}
depend on the chosen parameters for the simulations. Gradually increasing the 
gas sound speed (temperature) to an unphysical 30 \kms
(90.000 K) smoothes out the wiggles in the simulations.
As a result, a higher gas sound speed cancels out any change in 
the disk mass fraction (diamonds),
so that for $c_{\rm s} = 30$\,\kms the simulated gas velocity field is almost 
independent of $f_{\rm d}$. However, for smaller variations 
($\pm$5-10\kms) of the 
gas sound speed relative to our fiducial value of 10 \kms, the effect on 
the least squares analysis is weak. As for the effect of the
grid resolution, we studied it for the case of $f_{\rm d} = 60$ per cent 
(triangles in figure~\ref{chi}). The features resulting from the simulation 
on the 100 $\times$ 100 grid are smoother with respect to the fiducial
model, thus yielding a slightly worse agreement with the observed 
kinematics. On the other hand, the high resolution simulation 
(400 $\times$ 400 grid) yields an even worse agreement with
the observations. This can be understood by studying 
figures~\ref{shock_az_res} and~\ref{denscuts_all} 
which show that even though the 
amplitude and in many cases the peak positions of the wiggles 
on the 201 $\times$ 201
grid have nearly converged to their values on the 401 $\times$ 401 grid,
the shocks on the higher resolution grid have a different profile. They tend 
to be sharper and to have higher density and velocity contrast so that 
a measurement of the spatial overlap of the wiggles with observations finds 
significant discrepancies between the two models. Furthermore in 
comparison with the (lower resolution) observed gas kinematics the 
wiggles on the 401 $\times$ 401 grid exhibit
larger average deviations from the data points, resulting in a worse 
overall $\chi^2/$N value. Accordingly, we argue that 
simulations should be performed on a grid with a cell size comparable to
 the spatial resolution of the observational data that they will be 
compared to. Alternatively one can smooth high-resolution simulations to match
the spatial resolution of the observations. Finally, to explore the 
$\chi^2/$N values resulting from a different and more diffusive hydrocode, 
we plotted results from the beam 
scheme (filled circles) for simulations with fiducial values for 
the grid resolution, gas sound speed, and pattern speed.
Figure~\ref{chi} shows that the beam scheme 
seems to work relatively well 
for most of the disk regions. It should be noted however that the central 
region of the disk, where the beam scheme is the 
least successful, was excluded from the comparison.

Naively reading figure~\ref{chi}, we would conclude that the 45 
per cent disk model is the best match to the observations. 
However, in light of section~\ref{Parameters} which illustrated the
unlikelihood of simple 
simulations exactly matching every wiggle in the observations,
we make the following remarks on the $\chi^2$ analysis. 
If for a simulation performed with a given $f_{\rm d}$ there is a large
amplitude wiggle which is slightly spatially shifted from the wiggle in 
the observation, it will give a large deviation in the $\chi^2$ plot. As a 
result, the simulation may be discarded as one which gives a poor
match to the observation when in fact it could be that the simulation 
parameters were reasonable but not precise enough. For example, the 
physical sound speed at a wiggle location could have been very
different from the simulation sound speed, or the pattern speed which has 
the greatest influence on the positions of the wiggles, is not as constant 
as modeled. On the contrary, the simulations with smaller wiggles, caused 
either by low disk fraction, high sound speed, low resolution, or 
a very diffusive scheme, have the advantage that even if the peaks mismatch, 
they are small in the simulation so they do not carry as much weight in the 
calculation of the $\chi^2$ deviations. Hence these simulations
might give better $\chi^2$ matches. This suggests
that the $\chi^2$ analysis favors low $f_{\rm d}$ models.
This conclusion is not completely straightforward because
it is not true that all processes that weaken wiggles
improve the $\chi^2$ fit. For example, increasing the sound speed
from 10 to 30 km/s for the 20\% and 60\% disk fraction models
worsens the $\chi^2$.
However the robust trend that emerges from our analysis is that
changing the various modeling parameters for the higher $f_{\rm d}$ disks
causes larger variations in the $\chi^2$ than changing 
the same parameters for the lower $f_{\rm d}$ models.
In other words, 
getting the parameters ``right'' for the high $f_{\rm d}$ models is 
more crucial, than getting them ``right'' for the low $f_{\rm d}$ 
models. In that sense, the $\chi^2$ analysis favors the low $f_{\rm d}$ 
models. Given all the uncertainties in the modelling, variations
of the $\chi^2$ which are smaller than 10 or 20 per cent should not
be taken too seriously. Thus according to figure~\ref{chi}, the range 
for $f_{\rm d}$ lies between 0.2 and 0.85 but it is very difficult 
to distinguish models within this bracket. Unfortunately this range of
values for  $f_{\rm d}$ is 
very large since it allows for sub-maximal as well as marginally 
maximal disks. 

On the other hand, figure~\ref{wiggle_amplitude} only contains information
about wiggle amplitudes.  These are obtained by
subtracting an axisymmetric velocity field (a slowly varying function of 
$f_{\rm d}$, which explains
the different wiggle amplitudes for the measured data points) from 
both the simulation velocities and the observational velocities. 
Figure~\ref{wiggle_amplitude} corresponds to figure 8 in Paper I where 
the latter displays only results from the fiducial simulations.
For a fixed sound speed, the average wiggle amplitude scales almost
linearly with the disk mass fraction. As mentioned earlier, for a 
larger disk mass fraction, an increase of the gas sound speed smears 
out the wiggles thus decreasing their average amplitude.
Lowering the grid resolution (100 $\times$ 100) has the same effect, 
whereas increasing it (400 $\times$ 400) hardly makes any difference.
This is consistent with what we see in the $\chi^2$ plot 
(figure~\ref{chi}): the amplitudes of the wiggles in the 200 $\times$ 200
simulation are almost converged to their values on the 400 $\times$ 400
grid (see also figures~\ref{shock_az_res} and~\ref{denscuts_all}) thereby giving good agreement between
the two different resolution simulations in figure~\ref{wiggle_amplitude}.
As in figure~\ref{chi}, the beam scheme 
seems to work as well as the BGK scheme for the outer disk.
Neglecting error bars on the measurements, figure~\ref{wiggle_amplitude}
favors a $\sim$ 70 per cent disk model.
However, assuming an
error bar of 2 km/s on the average wiggle amplitude, the permitted
range for $f_{\rm d}$ according to figure~\ref{wiggle_amplitude} lies between
0.5 to 0.85. Taking the region of overlap of our two
methods, we find $f_{\rm d}$ between 0.5 and 
0.85. 
Even though both methods largely overlap in their predicted $f_{\rm d}$ values,
we argued that the $\chi^2$ criterion is not as well suited in excluding low
$f_{\rm d}$ values.
Therefore for a sample of galaxies, 
the second criterion is slightly better, since the associate errors are 
random, while the errors of the $\chi^2$ will always weigh in 
favor of small $f_d$ values.

Following an exploration of the simulation parameter space in this paper,
we understand why these global measurements do not give extremely accurate
measurements of the disk fraction. However, 
a robust conclusion of figures~\ref{chi} and~\ref{wiggle_amplitude} 
is that hydrodynamical simulations rule out a maximal disk model for 
NGC 4254, in agreement with our conclusions in Paper I and strongly
suggest a value of the disk fraction in the range 50 -- 85 
per cent.
Even when we make the gas in the disk (unreasonably) hot or when we use 
a very diffusive scheme, the maximal disk solution is never the best match 
to the observations. 
We conclude that the fiducial model used
in Paper I is indeed close to the best if not the
best estimate and that 
considering the simplified physics used in this analysis, our method 
puts surprisingly tight constraints on the amount of dark matter
present in high surface brightness galaxies.


\section{Conclusions} \label{conclusions}
We have demonstrated that despite simplifications in modelling
gas flows, hydrodynamical simulations can still
put strong constraints on the dark matter fraction of spiral galaxies.
Our main conclusions are:
\begin{itemize}
\item For the purposes of breaking the disk-halo degeneracy,
modeling gas flow across massive spiral arms may be preferable to
modeling the flow in the inner regions of strongly barred galaxies,
because gas flow in the inner region depends more critically on
the assumed sound speed of the ISM (see for example fig.~\ref{velcuts_cs})
and suffers more from numerical viscosity (fig.~\ref{difffd_codes_res}).

\item From the modelling parameters we considered, the pattern speed
is by far the most important parameter for determining the gas morphology.
This makes constraints on the pattern speed through comparison to the 
observed spiral morphology convincing.

\item  A detailed comparison between simulated and observed
velocity fields is challenging because of the dependence of the 
gas flow on many physical and numerical parameters. Nevertheless,
with a reasonable, physically motivated choice for the gas sound speed,
a grid resolution which is comparable to the resolution of the observations,
and a high-resolution hydrocode such as BGK, one can get a good estimate
of the pattern speed of the gravitational potential. Then, the baryonic
disk fraction remains as the primary parameter determining the flow.
While the technique discussed in this paper cannot yield an
exact number for the dark matter fraction in a spiral galaxy, it
can constrain it, conclusively ruling out certain values for $f_{\rm d}$.

\end{itemize}

\section*{Acknowledgments}
A. Slyz acknowledges the support of a Fellowship from the UK
Astrophysical Fluids Facility (UKAFF) where the computations reported
here were performed. The authors thank J.~Devriendt for a careful
reading of the manuscript and are grateful to the anonymous referee
for valuable comments.


\end{document}